\documentclass{aa}  

\usepackage{graphicx}
\usepackage{longtable}

\usepackage{booktabs}
\usepackage{txfonts}
\bibpunct{(}{)}{;}{a}{}{,} 
\usepackage{natbib}
\usepackage{caption}

\begin{document}

   \title{Physical parameters of IPHAS-selected classical Be stars.}

   \subtitle{I. Determination procedure and evaluation of the results.}
   
   \titlerunning{BCD classification of IPHAS CBe stars}

   \author{L. Gkouvelis
          \inst{1}\fnmsep\thanks{E-mail: leonardo.gkouvelis@uv.es}
          \
          , J. Fabregat\inst{1}
          \
          , J. Zorec\inst{2,3}
	\
	, D. Steeghs\inst{4}
       \
	, J. E. Drew\inst{5}
	\
	,  R. Raddi\inst{4}
	\
	, N. J. Wright\inst{5}
	\
	, J. J. Drake\inst{6}
          }
\authorrunning{L. Gkouvelis, J. Fabregat et al. 2016}

   \institute{Observatorio Astron\'omico, Universidad de Valencia, Catedr\'atico Jos\'e Beltr\'an 2, 46980 Paterna, Spain\
              \and   
             Institut d'Astrophysique de Paris, CNRS (UMR7095), 98 bis bd. Arago,
  75014 Paris, France\
             \and
             UPMC, Universit\'e Paris VI, (UMR7095), 98 bis bd. Arago, 75014 Paris,
  France\
             \and
               Department of Physics, University of Warwick, Gibbet Hill Road, Coventry CV4 7AL, U.K.\
             \and
              Centre  for Astrophysics Research, STRI, University of Hertfordshire, College Lane Campus, Hatfield, AL10 9AB, U.K.\
             \and
                Smithsonian Astrophysical Observatory, 60 Garden Street, Cambridge, MA 02138, USA\
                }

   \date{Received 2015; accepted 2016}

  \abstract
 {We present a semi-automatic procedure to obtain fundamental physical parameters and distances of classical Be (CBe) stars, based on the Barbier-Chalonge-Divan (BCD) spectrophotometric system. Our aim is to apply this procedure to a large sample of CBe stars detected by the IPHAS photometric survey, to determine their fundamental physical parameters and to explore their suitability as galactic structure tracers. In this paper we describe the methodology used and the validation of the procedure by comparing our results with those obtained from different independent astrophysical techniques for subsamples of stars in common with other studies. We also present a test case study of the galactic structure in the direction of the Perseus Galactic Arm, in order to compare our results with others recently obtained with different techniques and the same sample of stars. We did not find any significant clustering of stars at the expected positions of the Perseus and Outer Galactic Arms, in agreement with previous studies in the same area that we used for verification.}


 \maketitle 
 
\section{Introduction}

Classical Be (CBe) stars are main-sequence O, B and early A-type stars whose spectra show -or have shown- Balmer and other lines in emission. They are characterised by excess continuum emission in the ultraviolet, optical and infrared spectral ranges. Both the line and continuum emission arise from recombination processes in a hot, dense circumstellar decretion disk. The formation of these disks is not yet completely understood, although fast rotation, nonradial pulsations and magnetic fields are believed to be at their origin. A recent review of the nature of CBe stars and their main characteristics is presented by \citet{Rivinius2013}.\par

   \begin{figure*}
   \centering
 \includegraphics[trim={0.1cm 11.5cm 0.1cm 1.5cm},clip,width=\hsize]{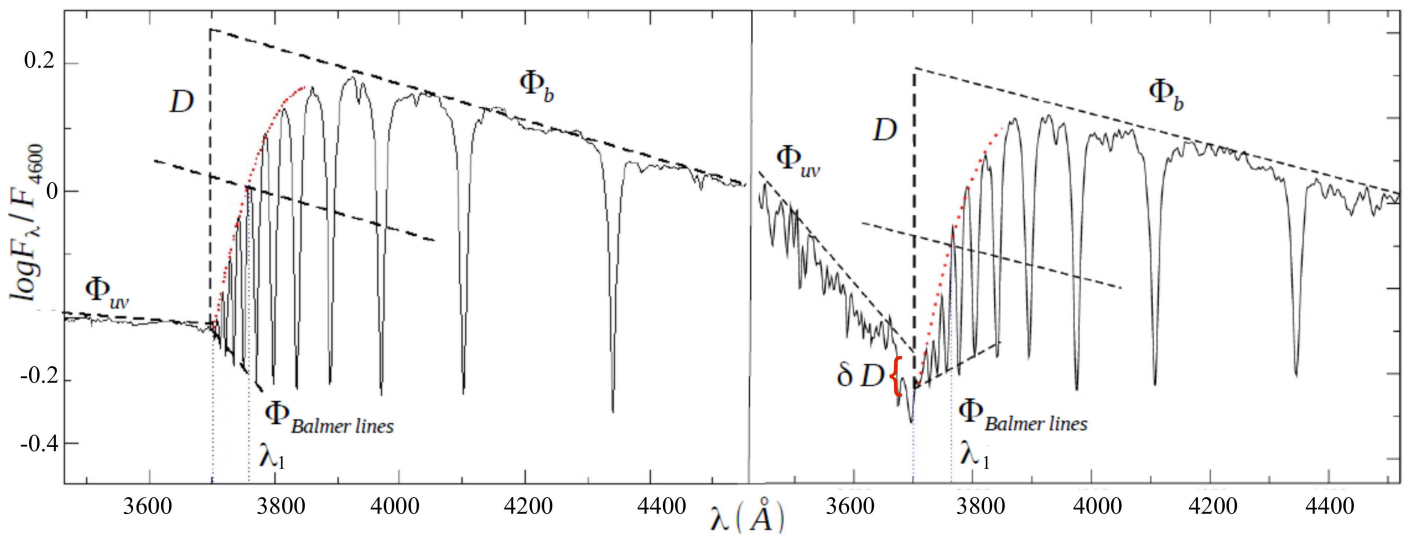}
   \caption{(a) Graphical explanation of the BCD $(D, \lambda_1,\phi_b,\phi_{uv})$ parameters in the spectrum of the B9.5V star HR4468; (b) determination of the BCD parameters of the emission line star IPHAS J002926.93+630450.2, with B9Ve spectral type. $\Phi_{Balmer~lines}$ is the line connecting the bottom of the higher Balmer lines, and $\delta D$ is the difference between the values of $D$ obtained from $\Phi_{Balmer~lines}$ and $\Phi_{uv}$. }
              \label{Fig:bcdparam}%
    \end{figure*}

In general CBe stars,  have had relatively little time to move far away from their birthplaces as they are short-lived objects - particularly those of the earlier types. Being main-sequence or slightly evolved stars, they are unlikely to be embedded in their parental clouds. In addition, they are intrinsically bright, with absolute magnitudes ranging from $\sim$0 to $\sim-4$, enabling their detection across the whole Galactic Plane, at least in regions not affected by heavy interstellar absorption. As such,  they are tracers relevant to the investigation of the spiral structure of the Galaxy.\par

The INT Photometric H$\alpha$ Survey of the Northern Galactic Plane \citep[IPHAS,][]{Drew2005,Barentsen14} is a 1800 deg$^2$ imaging survey covering the entire northern Milky Way at $|b| < 5^o$ in the $r$, $i$ and H$\alpha$ filters, using the Wide Field Camera (WFC) on the 2.5-metre \textit{Isaac Newton}  Telescope (INT) in La Palma. A first list of objects displaying H$\alpha$ emission was presented by \citet{witham2008}. Most of the objects from this list have been followed up spectroscopically, at $r < 17$ mag., at some point between 2005 and 2012, leading to a database of low resolution spectra of 2627 objects. Preliminary analysis reveals that about 70\% of this sample are CBe stars.\par

The general aim of this work is to obtain the fundamental physical parameters of the newly uncovered population of IPHAS CBe stars, which significantly increases the number of these objects known in the Galaxy. In addition, we aim to use this sample  to contribute to the investigation of the spiral structure of the Galaxy in the northern hemisphere, using them as tracers and the standard techniques of spectroscopic parallax to measure their distances. These sources will likely obtain moderately accurate trigonometric parallaxes from Gaia in the near future. Therefore, their proper characterisation would also aid their future exploitation for more detailed studies of Galactic structure with Gaia data.\par

To deal with such a large set of spectra we have developed a semi-automatic procedure to obtain the relevant physical parameters of the CBe stars, including the spectral type and luminosity class, effective temperature, interstellar extinction  and absolute magnitude. We have used the techniques and calibrations of the Barbier-Chalonge-Divan (BCD) spectrophotometric system \citep{Barbier1941,Chalonge1952}.\par

The aim of this paper is to describe the procedure we have developed to obtain CBe star astrophysical parameters from low resolution spectra, and to validate its results by comparing them with results obtained from different standard astrophysical techniques for samples of stars they have in common. A subsequent study, in preparation, will present the determination of the astrophysical parameters for the whole sample of the IPHAS CBe stars with available spectroscopy.\par

The paper is structured as follows: in Sect. 2 we describe the techniques of the BCD spectrophotometric system and their implementation in a semi-automatic procedure which allows us to analyse a large number of spectra efficiently,  with little human intervention; in Sect. 3 we describe the spectroscopic data used for this work; in Sect. 4 we present the work we have undertaken to validate our procedure, by comparing our results with results for two samples of stars obtained with different observational data and standard astrophysical techniques; in Sect. 5 we revisit the inferred spatial distribution of the CBe stars in the direction  of the Perseus Arm ($-1^{o}<b<4^{o}$,  $120^{o}<l<140^{o}$), and compare the results with recent studies in the literature.\par


\section{The BCD spectrophotometric system}

The BCD  spectrophotometric system was developed by \citet{Barbier1941}, and later by 
\citet{Chalonge1952}. The system is based upon
measurable parameters around the Balmer discontinuity (BD), in the  $3200-4600\ \AA$ spectral range. The basic
parameters that describe the energy distribution around the BD are: $D$, the Balmer jump
depth, given in dex, which is an effective temperature indicator; $\lambda_1$, the
position of the Balmer discontinuity, given in $\AA$ as the difference from
$3700\ \AA$, which is sensitive to the surface gravity; $\Phi_b$, a colour gradient which represent the slope of the Paschen 
continuum near the BD; and $\Phi_{uv}$, the slope of the Balmer continuum.\par

The BD is an easily visible spectral feature for stars ranging from early O to late F spectral
types. A modern description of the BCD system, together with a presentation of its advantages compared to other spectroscopic systems, is given by  \citet{Zorec2009}.\par

In Fig.~\ref{Fig:bcdparam} we present a graphical description of how the BCD parameters are measured. The value of $D$ is calculated at $\lambda = 3700\ \AA$, as $D=\log_{10}(F_{3700^+}/F_{3700^-})$,  where $F_{3700^+}$ is the flux of the extrapolated Paschen continuum and $F_{3700^-}$ the corresponding flux on the Balmer continuum.\par

To measure the $\lambda_1$ parameter, the mean spectral position of the BD is obtained as the intersection between the spectral continuum and the average of the extrapolated Paschen and Balmer continua at $3700\ \AA$. The line that represents the average of the $\Phi_{b}$  and $\Phi_{uv}$ fluxes is determined by the points $\log F_{\lambda} - D/2$, in the Paschen continuum region at $\lambda= 4000, 4150$ and  $4300\ \AA$ and $\log F_{3700} + D/2$ in the Balmer continuum at 3700\ \AA.\par

The parameters $D$ and $\lambda_1$ are independent of interstellar extinction, and allow the determination of the effective temperature and surface gravity by means of the calibration given by \citet{Zorec2009}. The third parameter $\Phi_{b}$, which measures the slope of the Paschen continuum, is a function of both the effective temperature and the interstellar extinction, and can be used to measure the interstellar reddening by means of the expression:
\begin{equation}
A_V = 3.1E(B-V) = 2.1(\Phi_{b} -\Phi_{b}^{0} ) 
\end{equation}
From \citep{Aidelman2012}, where $\Phi_{b}^{0}$ is the intrinsic gradient obtained from $D$ and $\lambda_1$ by means of the calibration given in \citet{Chalonge1973}.\par

 \begin{figure}
   \centering
   \includegraphics[width=\hsize]{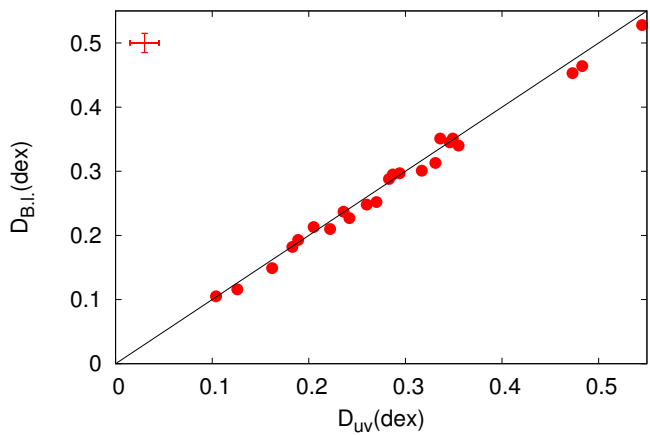}
      \caption{Comparison of the Balmer jump depth calculated for 23 absorption-line B-type stars, using both the ultraviolet continuum and the Balmer lines limit as explained in the text. At the left upper part it is shown the typical error in the determination of  $D$, 0.015 dex.}
         \label{FigStandard}
   \end{figure}

\begin{figure}
   \centering
   \includegraphics[width=\hsize]{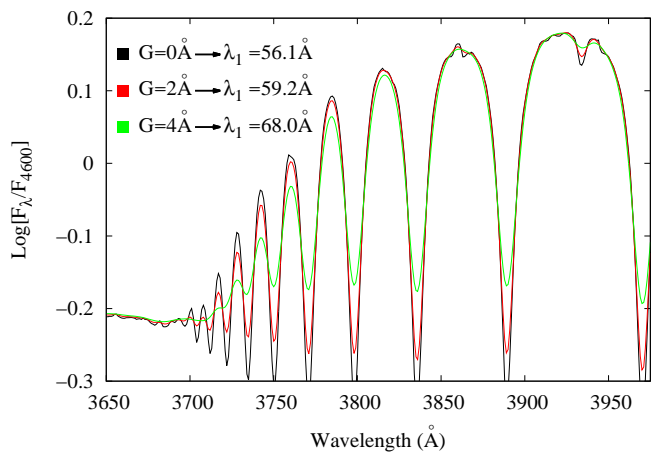}
      \caption{The region around the BD area of the standard B-type star HR4468. The black line represents the original spectrum, and red and green lines the spectrum convolved with Gaussian filters of $2\AA$ and $4\AA$ width respectively. }
         \label{Fig:convolution}
   \end{figure}

\subsection{The determination of the photospheric Balmer discontinuity of Be stars}

The procedure described above to determine the Balmer discontinuity depth is, however, in general not applicable to CBe stars. Due to the lower pressure of the circumstellar disc, a second Balmer jump exists for these objects in the ultraviolet region of the spectrum, which is either in emission or absorption \citep{Kaiser1989}. In the case of emission, we extract the flux at $F_{3700^-}$ as the point where the bottom of the higher order Balmer line series merges into the continuum.

In Fig.~\ref{Fig:bcdparam}b we illustrate this last procedure. To obtain the flux at $F_{3700^-}$, instead of using the extrapolated Balmer continuum $\Phi_{uv}$, we extrapolate the line traced through the bottom of the higher Balmer lines, which is labeled $\Phi_{Balmer~lines}$. In the case of non emission-line stars, both methods lead to the same $D$ values, as shown in Figs.~\ref{Fig:bcdparam}a and \ref{FigStandard}, and in columns 6 and 7 of Table~\ref{table:2}. The mean of the residuals between the two different measurement procedures is 0.010$\pm$0.007 dex, significantly smaller than the characteristic error of 0.015 dex associated with the measure of the $D$ parameter in the BCD system.\par

In the case of CBe stars, the value of $D$ obtained in this way is the true photospheric value, not affected by circumstellar emission or absorption in the Balmer and Paschen continua. The validity of this statement is fundamental for the applicability of the BCD system to the determination of the physical parameters of CBe stars, and hence in Appendix A we discuss this issue in detail.

\subsection{Setting the spectral resolution}

The original calibration of the stellar parameters in the BCD system was made empirically from spectra with a mean resolution of 
$\Delta \lambda  \sim 8\ \AA$  at the BD. Although the $D$ parameter is independent of the resolution, the mean position of the BD, and hence the $\lambda_1$ parameter, varies with the spectral resolution. This effect is illustrated in Fig.~\ref{Fig:convolution}, where we present the spectrum of the B-type star HR4486 at different resolutions.\par

Because the determination of astrophysical fundamental parameters proceeds with calibrations obtained with the original BCD parameters, to measure $\lambda_1$ we have to reduce the resolution of our spectra to the characteristic resolution of the original BCD spectrophotometric system. This is done by convolving the spectra with a Gaussian filter of the appropriate width, so that the resolution to apply the BCD formalism is obtained by
\begin{equation}
R_F = \sqrt{R_I^2 + G^2}
\end{equation}
where $R_F$ is the resolution required for the BCD method, $R_I$ is the initial resolution of the spectra, and $G$ is the width of the Gaussian filter.\par

 Setting of the correct spectral resolution is a complex issue, and depends on the available set of data. In Sect. 3.3. we explain how we have selected the appropriate correction of the resolution for the spectra analysed in this work.  

\subsection{Absolute magnitudes and distances}

Absolute magnitudes in the Johnson $V$ band (M$_V$) are directly obtained from the $D$ and $\lambda_1$ parameters by means of the calibration given in Table 3 of \citet{Zorec1991}. They are converted to absolute IPHAS $r$ magnitudes using the intrinsic $(V-R_C)$ colours for dwarfs and giants supplied by Fabregat (in preparation), and assuming that Cousins $R_C$ and IPHAS $r$ magnitudes in the Vega system are identical within the errors involved in our procedure.\par

Distances are obtained by means of the standard spectroscopic parallax techniques, by comparing the absolute M$_r$ magnitudes with the observed $r$ magnitudes supplied in the IPHAS Second Data Release \citep[IPHAS DR2,][]{Barentsen14}, corrected for the interstellar absorption and the circumstellar excess due to the added flux of the disk emission.\par

Intrinsic $r$ magnitudes corrected for interstellar absorption were computed using the relationship A$_r$ = 0.84 A$_V$, from \citet{Fiorucci03}. To correct for CBe circumstellar continuum emission we followed the method described in Sect. 3.3 of \citet{Raddi2013}, which follows the earlier work by  \citet{Dachs1988} that investigated the correlation between  $EW(H\alpha)$ and the circumstellar colour excess  $E^{cs}(B-V)$.  The relations adopted in this work are 
\begin{equation}
E^{cs}(B-V) \approx 0.02\frac{EW(H\alpha)}{-10\AA},
\end{equation}
and
\begin{equation}
f_D = \frac {F_D} {F_D + F^*} \approx 0.1 \frac {EW(H\alpha)} {-30\AA},
\end{equation}
From these values we obtain the circumstellar emission in the $r$ band, $\Delta r$, by interpolating from Table 5 of
 \citet{Raddi2013}.  The $\Delta r $ correction applied  for each source is given in Table~\ref{table:results2}.\par

\subsection{Code principles}

In order to analyse a large sample of spectra in a reasonable amount of time we have developed  a semi-automatic procedure, only requiring user interaction to evaluate output diagrams at some important steps during the treatment of each spectrum. The basic inputs of the programme are the observed spectra, with the wavelength scale in $\AA$ and the flux in $erg s^{-1} cm^{-2}$, and a list with the names of the files containing each spectrum.\par

The pipeline used to treat each spectrum requires a series of inputs that the user provides as the analysis proceeds. The first step of the procedure is to ask the user for the dispersion of the spectra, and for the width $G$, in $\AA$, of the Gaussian filter that the spectra require in order to transform their resolution to the original resolution of the BCD system. This first step is valid for the whole run and applies to all the spectra in the list.  The next step is the analysis of the spectra one by one. As each spectrum is presented for analysis, it is first thoroughly checked to establish its wavelength range and to detect possible recording flaws.\par

One of the steps which requires user interaction is the wavelength scale, whose possible variations, caused either by errors in the wavelength calibration or by the radial velocity of the star, may affect the position of the Balmer discontinuity and the determination of BCD parameters. The largest shifts in the wavelength calibration were found to range from 3 to 10$\ \AA$.\par

A flux normalization is applied at $\lambda\  4600\ \AA$, and the spectral coordinates transformed to $1/\lambda-\log(F/F_{4600})$, to measure the gradient $\Phi_b$ consistently with the definition of colour gradient \citep{Allen}. This also enables a first estimate of the BD depth, $D_1$ (dex), which is considered to be preliminary because the Paschen continuum cannot be well represented by a straight line in these coordinates. $D_1$ is used to derive an approximate value of the stellar effective temperature. The spectrum is subsequently divided by a Planck function at this temperature. The Paschen continuum in the normalised, divided spectrum closely approach to a straight line, and can be extrapolated without ambiguity. The determination of $D$ as the ratio of fluxes at  $3700\ \AA$ removes the contribution of the Plank function. 

To measure the mean spectral position of the BD we calculate the line determined by the points $\log F_{\lambda} - D/2$ at $\lambda= 3700, 4000, 4150$ and  $4300\ \AA$, where $F_{\lambda}$ is the flux of the extrapolated Paschen continuum. We characterise the spectral continuum as a parabolic fit to the pseudo-continuum in the $3700 - 3850\ \AA$ region. The intersection between these two lines determines the value of the $\lambda_1$ parameter.\par
  
\begin{figure}
   \centering
   \includegraphics[width=\hsize]{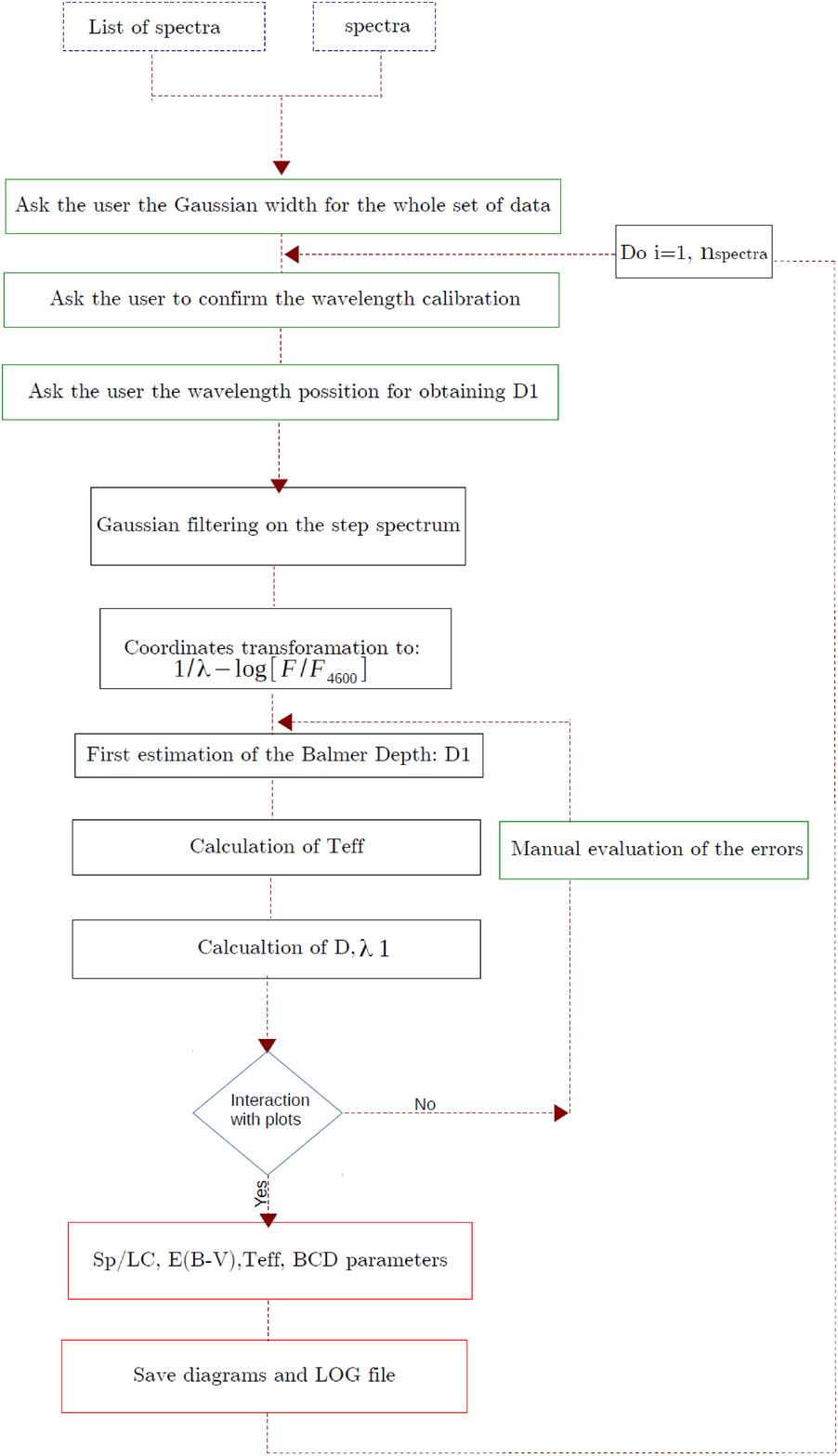}
      \caption{Simplified block diagram of the spectral analysis pipeline based on the BCD classification system presented in this work.}
         \label{Fig:BlockD}
   \end{figure}

 For each spectrum, a series of diagrams are constructed, with which the user
can interact to control the progress of the analysis. The steps in the algorithm are shown in Fig.~\ref{Fig:BlockD}.\par
The output of the program is a log file containing the parameters calculated for each spectrum:  name, BCD$(D, \lambda_1)$, $T_{\rm eff}$, spectral type and luminosity class, $\Phi_b$,  $\Phi_{b}^{0}$, $E(B-V)$ and M$_V$.  

\section{Observations and selection of the data.}

\subsection{IPHAS bright sample with FLWO/FAST}
Follow-up spectroscopy of the emission line objects photometrically detected by IPHAS \citep{witham2008}  was performed during the years 2005 to 2012 at the 1.5m Fred Laurence Whipple Observatory (FLWO) Tillinghast telescope on Mount Hopkins in Arizona, using the FAST spectrograph \citep{Fabricant1998}. The data were taken with the 300 lines $mm^{-1}$ grating, and a projected slit width of 3''. The data span a wavelength range from 3500 to $7500\ \AA$ at a spectral resolution of  $\Delta\lambda\simeq6\ \AA$.\par

The data were processed at the Telescope Data Center at the Smithsonian Astrophysical Observatory. The spectra were delivered without flux calibration. As  explained in Sect. 2, the BCD method requires at least an accurate relative flux calibration, especially in the blue and near ultraviolet part of the spectrum, around the Balmer discontinuity. This spectral region is very hard to observe in a flux-calibrated way, due to the weakness of incandescent flat field calibration lamps, CCD efficiency and optical coating properties. For this reason we took special care in performing the flux calibration.\par

For each different night we selected calibration spectra from the FAST archive, to ensure that all spectra were calibrated with flux standards observed the same night. The calibration has been done using standard image reduction and analysis facility (IRAF) routines.\par 

For 31 stars we have two spectra obtained at different epochs, and for one more three spectra. {\bf  In Appendix B we present overplots of the different spectra obtained for each star and a detailed discussion on the reliability of the relative flux calibrations and the spectral parameters derived from them}.\par

We estimate the mean error of the flux calibration as the difference between the individual values of the flux divided by the mean value. The mean error in the flux of the calibrated and normalised spectra amounts to 5.1$\pm$4.5\% at $4000\ \AA$, and to 15$\pm$14\% at $3600\ \AA$. The last value in the ultraviolet continuum is significantly larger, as expected. However, as explained in Sect. 2.1, to obtain the lower limit of the Balmer discontinuity of the CBe stars we don't use the extrapolated ultraviolet continuum, but instead the bottom of the higher Balmer lines. Hence, even large errors in the flux calibration short of the Balmer discontinuity, which eventually could lead to an inaccurate determination of the slope or the position of the Balmer continuum, will not have any impact in the determination of the CBe stars physical parameters used through this work.\par

 The mean difference between the measured $D$ and $\lambda_1$ parameters from different spectra of the same star are 0.032 dex and $7.3 \AA$  respectively. These differences are of the same order as the ranges in $D$ and $\lambda_1$ spanned by one spectral subtype and one luminosity class respectively, as can be seen in Fig. 10 of \citet{Zorec2009}.\par
 
For this work we selected the sources with spectral features characteristic of OB-type stars. Among them  we selected the spectra which have S/N $\geq 30$ around the Balmer discontinuity ($\sim3700\ \AA$), where the BCD parameters are measured.  Finally, we rejected the spectra in which the continuum around the BD is not well defined. This rejection was applied automatically each time that the $\chi^2$ of a parabolic fit to the pseudo-continuum drawn in the BD region is larger than $3\sigma$ as compared to the average of fits obtained for the entire stellar sample. We considered that in those cases we could not derive a reliable $\lambda_1$ parameter.\par

We examined an initial sample sample of 612 FAST spectra in the Perseus Arm region. After rejecting the spectra whose characteristics did not correspond to an OB-type star, and those not meeting the criteria presented in the previous paragraph, we kept a final sample of 257 spectra for further analysis.\par 

\subsection{INT and NOT spectra}

Mid-resolution (2-4\ \AA)  and high S/N (30-100 at 3700\ \AA) spectra of 67 CBe stars were obtained at the Roque de los Muchachos Observatory in La Palma, Canary Islands, Spain. The telescopes and instruments used were the Isaac Newton Telescope (INT) equipped with the Intermediate Dispersion Spectrograph (IDS), and the Nordic Optical Telescope (NOT), using the Andaluc\'\i a Faint Object Spectrograph and Camera (ALFOSC). A sample of spectrophotometric standard stars, for relative flux calibration, and MK standard stars, were also observed. A complete description of this data sample is given in Sect. 2.3 of \citet{Raddi2013}.\par

These spectra have already been analysed by \citet{Raddi2013}. In this work we will use them in the procedure to determine the value of $G$, the width of the gaussian filter required for the  FLWO/FAST spectra, as described in the next subsection, and to compare the results of our analysis with those obtained by \citet{Raddi2013} via energy distribution fitting to appropriate model atmospheres.\par

\subsection{Gaussian filtering}
For every dataset, with different resolution, we must find the appropriate gaussian width with which to convolve the spectra in order to achieve the desired BCD resolution. The best way of doing this is to observe BCD standard stars with the same instrumental configuration as the programme stars. A list of BCD standards, that is, a  sample of stars for which standard values of the $D$ and $\lambda_1$ parameters are know, is given by \citet{Zorec1991}.\par

\begin{figure}
   \centering
   \includegraphics[width=\hsize]{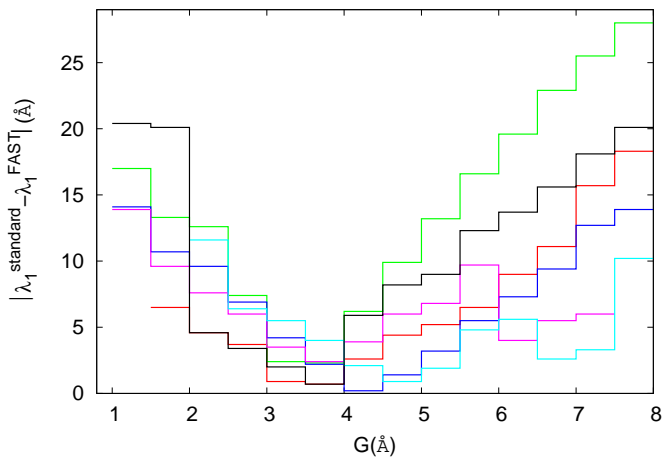}
      \caption{Differences in the $\lambda_1$ parameter obtained with the INT/NOT spectra and the FAST spectra convolved with different Gaussian widths ranging from 1.5 to 8$\ \AA$, at steps of 0.5$\ \AA$, for a subset of the stars in common between the two samples. Each colour represents a spectrum and the minimum of the $\Delta \lambda_1$ difference corresponds to the convolution that best fits the INT/NOT spectrum. }
         \label{Fig:Gaussian}
   \end{figure}

No BCD standards were observed with the FAST spectrograph, and only two were obtained at the INT and two more at the NOT. In order to determine the value of $G$ (Eq. 2) and correct the resolution for all spectra we take the following steps:

\begin{enumerate}

\item   As a starting point we used the INT and NOT spectra of the stars for which the standard $D$ and $\lambda_1$ are known. For each star we computed the BCD parameters using Gaussian filters of different width, and compared the obtained values with the standard ones. The best agreement for each star was obtained with the Gaussian widths presented in Table~\ref{table:1}.\par

\item We obtained the BCD parameters of all MK standard stars observed with the INT and NOT, using the widths in Table~\ref{table:1}. From the BCD parameters we obtained the spectral types and luminosity classes, and compared them with the standard MK ones. This comparison is presented in Table~\ref{table:2}. The agreement between the MK and BCD classification is fairly good, within two spectral subtypes and one luminosity class for most of the stars. When comparing the classification of stars in both systems, we have to keep in mind that the MK system assigns discrete spectral types, where each encompasses comparatively large intervals of ($D, \lambda_1$) parameters. Obvious differences can then appear when interpreting continuous runs of BCD parameters with discrete MK assignments.\par

From the above procedure, we find $G =$ 4.8, 4.0 and 3.5$\ \AA$ as the widths of the Gaussian filters to reduce the  resolution of the INT (with two different instrumental configurations) and NOT spectra, respectively. We subsequently applied Gaussian filters of these widths to the whole sample of INT and NOT spectra, and computed the BCD parameters for all of them.\par

\item The final step is to set the resolution for the FAST spectra. The procedure is described graphically in Fig.~\ref{Fig:Gaussian}. We convolved the FAST spectra, one by one, starting from a Gaussian width  of 1.5$\ \AA$   up to 8$\ \AA$, with a 0.5$\ \AA$ step. For all the stars in common between the INT/NOT and FAST samples we plot the difference  $\mid\lambda_{1,INT/NOT}-\lambda_{1,FAST}\mid$ versus the Gaussian width. The minimum difference is found at $G=4.25\ \AA$, and we assume this value to be the Gaussian width, $G$, with which to convolve all the FAST spectra prior to the determination of the BCD parameters from them.\par

\end{enumerate}

\begin{table*}
\caption{BCD standard stars observed with the INT and NOT telescopes. Columns 4, 5 and 6 are the standard BCD spectral type and $D$ and $ \lambda_1$ parameters respectively, from \citet{Zorec1991}. The different $ \Delta \lambda$ and dispersion values for the two stars observed at the INT are due to different instrumental configurations. The last column lists the width of the Gaussian filter used to reproduce the standard BCD parameters from each spectra.} 
\label{table:1}      
\centering          
\begin{tabular}{c c c c c c c c c}     
\hline\hline       
             NAME    & Telescope    & Sp/LC$_{\rm MK}$   &   Sp/LC$_{\rm BCD}$    &    $ \lambda_1(\AA)$   &     $D(dex)$  &$ \Delta \lambda(\AA)$&Disp.$(\AA/pix)$&$ G(\AA)$ \\
\hline 
              HR 533   &    INT &     B2 V  &         B2 V   &       65 &          0.144 &         4                     &   1.85     &  4.8  \\
              HR 1122  &   NOT &    B5 III  &        B5-6 III   &   42 &          0.286 &        2                     & 0.72   &   3.5    \\
              HR 2347  &   NOT &    B9 V  &        B9 V   &        64 &         0.422 &         2                     &   0.72   &  3.5   \\
              HR 2461  &   INT &      B8 III &        B7 III   &       41 &          0.340 &         3                      &   1.40   &  4.0   \\

\hline                  
\end{tabular}
\end{table*}

\begin{table*}
\caption{Sample of 23 MK standard stars observed with the  INT and NOT telescopes, covering almost all the B sub-types. We present the measured BCD parameters and the comparison between the standard MK type and the obtained BCD spectral classification.  }             
\label{table:2}      
\centering          
\begin{tabular}{c c c c c c c c }     
\hline\hline       
             NAME   & Telescope     & Sp/LC$_{\rm MK}$   & Sp/LC$_{\rm BCD} $  & $\lambda_1(\AA)$   &  $ D_{uv}(dex)$  &     $D_{B.l.}(dex)$   &    $ \mid D_{uv}- D_{B.l.}\mid$  \\
\hline 

     HR 533 & INT    & B2 V & B2 V &79.4         & 0.162 & 0.149 &  0.013 \\
       HR 927 & INT &    B8 V  & B7-8 V & 54.3  & 0.346 &  0.345 &  0.001 \\
       HR 1122 & NOT  &   B5 III & B5-6 III & 36.1&  0.236 & 0.237   &  0.001 \\
        HR 1399  &INT&   B5-6 V  &B6 V&  68.5  &  0.287   &    0.295 & 0.008       \\
    HR 1497 &INT&   B3 V  & B7-8 V& 77.0 & 0.294      &    0.297 & 0.003     \\
     HR 1576 &INT&   B9 V  & B7 V&    67.9&  0.283      &    0.288 & 0.005     \\
      HR 1595 &NOT&  B2 V  & B1 V&  63.6  &    0.126   &    0.116 & 0.010       \\
         HR 1760 & INT& A3V & >A2         & 74.9 &   0.545 & 0.528 &   0.017 \\
   HR 1808 &NOT & B5 V & B4 V       &  62.1 &   0.260 & 0.248 &  0.012  \\
  HR 1820  & NOT & B2 V & B3 V    & 60.3  &  0.189&    0.193   &  0.004  \\
  HR 1860  &NOT&   B6 V  & B6 V& 60.5  &  0.317   &    0.301 & 0.016       \\
  HR 1863  &NOT&   B2.5 V  &B4 V& 69.0  &   0.222   &    0.210 & 0.012       \\
  HR 1892 & NOT & B1 V & B0 V     & 61.4  & 0.104 &  0.105   &  0.001 \\   
   HR 2010 & NOT & B9 IV & A0-1 V &  65.7  &  0.483  & 0.464   &  0.019 \\  
  HR 2116 &NOT&  B8 V  & B7-8 V&  62.7  &    0.355   &    0.340 & 0.015       \\
   HR 2161 & NOT& B3 V &B3 V   &  64.4 &    0.205  &   0.213  &  0.008 \\
 HR 2344  &   INT & B2 V  &  B3 IV & 54.0  &  0.183  &  0.182   & 0.001 \\
HR 2347   &  NOT & B9 V  &  A0 V  & 66.6&   0.473  &  0.453   &  0.020 \\
HR 2461   &  INT  &  B8 III&  B7-8 IV  &45.4 &   0.349 & 0.351  & 0.002 \\
  HR 2490 &NOT &  B3 IV  &  B3 IV& 51.6 &     0.242   &    0.227 & 0.015       \\
 HR 2840 &  INT   & B6 IV  & B7 IV   &47.7&    0.336   & 0.351   & 0.015 \\
HR 7996 & INT    &  B3 III  & B3-4 IV  & 48.9&   0.270  & 0.252  & 0.018 \\
HR 8403 & INT    & B5 III&  B5 V   & 54.3&   0.331  &    0.313  &  0.018 \\

\hline                  
\end{tabular}
\end{table*}

\section{Evaluation of the procedure}

We now validate the described procedure, by comparing our results with results from the literature obtained for the same stars with different astrophysical techniques. For this comparison, we used the results obtained from spectroscopic data analysis by  \citet{Raddi2013}, and from Str\"omgren photometry calibration of data presented by \citet{Fabregat2005} and \citet{M2013}.\par

\subsection{Validation against previous spectroscopic analyses}

\citet{Raddi2013} studied a group of 67 candidate CBe stars in the region of the Perseus Arm, by analysing the mid resolution spectra obtained with the INT and NOT telescopes at La Palma described in Sect. 3.2. They determined their spectral types and measured their colour excess via spectral energy distribution fitting to appropriate model atmospheres in the blue part of the spectrum (3800-5000 \AA ).\par 

For this comparison we use 35 spectra that meet the selection criteria described in Sect. 3.1. The results are presented in Table~\ref{table:3}. To provide an additional element of comparison, in Table~\ref{table:3}  we also present spectral types and luminosity classes obtained from the INT/NOT spectra by means of the standard MK classification procedures, using only the strength and widths of the spectral lines, and the ratios between lines, such as temperature and luminosity criteria, as described for example by \citet{GrayCorbally}. In the last two columns we present the $E(B-V)$ obtained from the measured BCD parameters by means of Eq.1, and the absolute magnitudes in the scale defined by \citet{Zorec1991}. We estimate a mean error of 0.05 mag. for the $E(B-V)$ determination, obtained from the standard propagation of the errors through Eq. 1, considering the characteristic uncertainties of the BCD parameters involved. The mean uncertainty of the $M_V $ values amounts to 0.15 mag. \citep{Zorec1991}.\par

From Table~\ref{table:3} we can see a general good agreement within the three classification systems. When comparing our results with those of \citet{Raddi2013} we find  that 29 of the 35 stars agree within one or two spectral subtypes. In the remaining six stars, however, there are large differences. Better agreement, with differences not larger than two sub-spectral types, is found between the BCD results and the MK classification applied to the same spectra.\par

\begin{figure}
   \centering
   \includegraphics[width=\hsize]{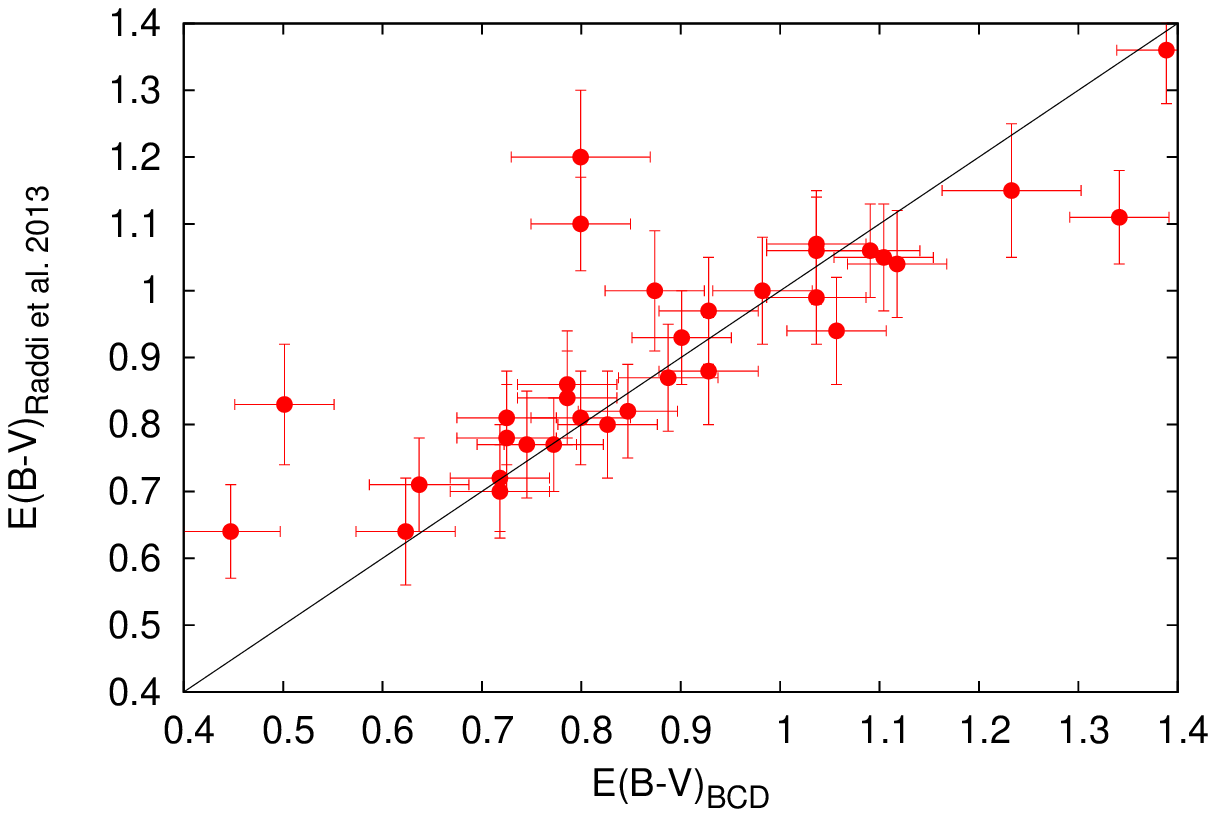}
      \caption{ Comparison between the colour excesses obtained from the FAST spectra with the procedure described in this work and the values given by \citet{Raddi2013}. }
         \label{Fig:EvsE}
   \end{figure}

In Fig.~\ref{Fig:EvsE} we present the comparison between the $E(B-V)$ obtained by \citet{Raddi2013} and our procedure. A mean error of 0.05 mag. for our determination of the $E(B-V)$ is assumed, as stated above. For the values of \citet{Raddi2013} we used the errors quoted by these authors. There is a good agreement between the two sets of data, with a mean difference of 0.04$\pm$0.15 mag.\par

\begin{table*}
\caption{BCD parameters and classification of 35 IPHAS CBe stars with INT/NOT spectroscopy, and comparison with MK classification and spectral classification given by \citet{Raddi2013}.  }             
\label{table:3}      
\centering          
\begin{tabular}{c c c c c c c c c r }     
\hline\hline       
             NAME        & D (dex)   &$ \lambda_1 (\AA)$  & $\Phi_b $ &   $\Phi_{b}^0 $& Sp/LC$_{\rm BCD}$  &Sp/LC$_{\rm MK}$& Raddi et al.& $E(B-V)$ & $M_V $  \\
\hline 

$J002441.73+642137.5 $  & 0.291    & 36.5    &    2.85	 &   0.80 &   B5-6III &   B5III       &  B5III     &     1.38	  & -1.9    \\
$J002926.93+630450.2 $  & 0.355    & 72.3    &    1.51	 &   0.85 &   B9V     &     B5-8V  &  B7V     &     0.44 	&   1.0     \\
$J004014.89+651644.0 $  & 0.174    & 79.4    &    2.26	 &   0.73 &     B3V   &   B2-3V   &  B2V     &     1.03     &    -1.5	  \\
$J005029.25+653330.8 $  & 0.280    & 53.5    &    2.41	 &   0.78 &   B5-6V &    B3-4V   & B7IV     &	   1.10 &	 -0.8	 \\
$J005436.84+630549.9 $  & 0.195    & 63.3    &    2.10	 &   0.73 &    B3V    & B3V          & B2-3V &      0.92     &	-1.4\\
$J005611.62+630350.5 $  & 0.280    & 69.9    &    1.53	 &   0.79 &   B7V     &    B5-6V   &  B5V   &      0.50       &	-0.1\\
$J005619.50+625824.0 $  & 0.288    & 70.2    &    1.07	 &   0.80 &    B7V    &   B3-5V    &  B5V   &	    0.18       &	      -0.1   \\
$J010707.68+625117.0 $  & 0.291    & 60.8    &    2.40	 &   0.79 &    B5V    & B6-7Ia     &  B5V    &	     1.09      &	-0.1	  \\   
$J012405.42+660059.9 $  & 0.246    & 54.8    &    1.94	 &   0.76 &    B4V    &   B2-3V    &   B6IV  &    0.79      &	  -1.0         \\
$J012751.29+655104.0 $  & 0.201    & 63.5    &    2.02	 &   0.73 &     B3V   &   B3V      &   B7V   &    	  0.87     &      -1.3	  \\
$J014458.14+633244.0 $  & 0.346    & 63.9    &    1.92       &   0.82 &  B7-8V  & B3-4V      &  B7IV        &  	 0.74     &	     0.6      \\
$J014620.44+644802.5 $  & 0.315    & 54.8    &    2.02	 &   0.80 &    B6V    &  B5V        &   B7V           &  	  0.82    &	-0.4	  \\
$J014905.18+624912.3 $  & 0.111    & 34.5    &    2.98       &   0.70 &   B2Ib      & B2IV-III	&  B3IV  &	   1.54     &	 -5.6\\
$J015037.67+644446.9 $  & 0.203    & 64.1    &    1.87       &   0.73 &   B3-4V  &  B3-5V    &  B4V        &  	 0.77      &	    -1.1       \\
$J015246.27+630315.0 $  & 0.403    & 55.9    &    2.02       &   0.86 &   B9V      &  B8-9V      &   B8-9III      &	 0.78  &	0.1	     \\  
$J015613.22+635623.8 $  & 0.165    & 68.8    &    1.64       &   0.72 &   B2V      &   B2V 	 &   B3V       &      0.62     &   -2.0  \\
$J015918.32+654955.8 $  & 0.321    & 69.0    &    2.22       &   0.85 &   B8V      &   B7-8V    &  B6IV        &  	0.92        &	 0.5	    \\
$J015922.53+635829.3 $  & 0.234    & 57.1    &    1.91	  &   0.75 &   B4V      &   B2-3V     &   B2-3V   &   	 0.78   &	  -0.9\\
$J022337.05+601602.8 $  & 0.185    & 52.9    &    2.26        &   0.73 &   B3-4IV  &   B3-4V     &   B7IV     &  	 1.03    &	      -2.2\\
$J023031.39+594127.1 $  & 0.268    & 50.0    &    1.95        &   0.77 &   B5-6V   &  B6-7V      &   B9V      &	  0.79   &	    -1.0\\
$J023404.70+605914.4 $  & 0.184    & 60.5    &     2.71	   &   0.73  &   B3-4V   &    B2V	        &  B3IV      &        1.34 &	      -1.5\\
$J023642.66+614714.9 $  & 0.209    & 47.2    &    1.82	   &   0.75 &   B3-4IV  &    B1-3V    &   B5V    &		 0.72   &	  -2.0\\
$J023744.52+605352.8 $  & 0.142    & 43.6    &    2.52        &   0.70 &   B2III      &    - 	               &   B8    &  	     1.23        &	    -3.7\\
$J024054.96+630009.7 $  & 0.214    & 50.3    &    1.99	   &   0.74 &   B3-4IV  &  B1-3V	      &  B6      &	    0.84    &	  -1.6\\
$J024146.74+602532.2 $  & 0.176    & 52.4    &    2.37        &   0.72 &   B2IV      &  B3III         &  B7V        &		1.11    &	      -2.2\\
$J024159.21+600106.0 $  & 0.197    & 57.6    &    2.18        &   0.73 &   B3-4V   &  B5V          &   B5V    & 	     0.98       &	     -1.6\\
$J024317.68+603205.5 $  & 0.120    & 44.2    &    2.26        &   0.70 &   B1III      &  B1.5-2III   &  B7V       &	  1.05    & 	-4.3\\
$J024504.86+612502.0 $  & 0.257    & 43.4    &    2.09        &   0.78 &   B5-6IV  &   B6   	       &    B7IV     &   0.88      &	   -1.8\\
$J024618.12+613514.7 $  & 0.171    & 59.8    &    1.79	   &   0.72 &   B2V       &   B2V           &    B3V     &    0.72   &	   -1.9\\
$J025016.66+624435.6 $  & 0.345    & 44.4    &    1.77        &   0.83 &   B7-8IV  &   B8-9V     &    B8-9III  &		0.63     &	     -0.9\\
$J025059.14+615648.7 $  & 0.147    & 62.7    &    1.76	   &   0.70 &   B2V       &   B3V          &   B3-4V    &       0.71  &	     -1.9\\
$J025233.25+615902.2 $  & 0.244    & 40.7    &    1.83	   &   0.77 &   B5-6III  &  B4V          &    B7V      &	   0.71      &	       -2.1\\
$J025448.85+605832.1 $  & 0.296    & 59.5    &    1.97	   &   0.79 &   B5-6V   &  B6-7V	      &  B6V	     &     0.79 &	    -0.2\\
$J025610.40+580629.6 $  & 0.161    & 55.2    &    2.04	   &   0.71 &   B2V       &   B3-4V     &   B5V     &  	     0.90   &	    -2.4\\
$J025700.49+575742.8 $  & 0.152    & 47.1    &    2.24	   &   0.71 &   B2III      &    B3-4V    &   B4V      &  	1.03    &         -3.2\\
\hline       
\end{tabular}
\end{table*}

\subsection{Validation with $uvby\beta$ photometry}

\citet{M2013} present a catalogue of $uvby\beta$ Str\"omgren-Crawford photometry for 35974 stars in the galactic anticentre direction. Thirteen classical Be stars for which we have FLWO/FAST spectroscopy have photometric data in this catalogue. In addition, two more CBe stars  in the area of the galactic open cluster NGC 663, with available spectroscopy, have $uvby\beta$ photometry published by \citet{Fabregat2005}.\par

For these stars we have obtained the interstellar reddening, spectral classification and absolute magnitude, using the $uvby\beta$ photometry calibrations given by  \citet{crawford78},  \citet{Balona1984} and \citet{Moon1986}. We have followed the procedures described by \citet{Fabregat1998} to apply the above calibrations to CBe stars, for which the circumstellar emission in the Balmer and Paschen  continua and in the H$\beta$ line have large effects on the photometric indices.\par

In Table~\ref{table:4} we compare the results from the $uvby\beta$ photometry analysis with those obtained with the BCD techniques described in the previous sections. The differences in the spectral types between the two methods is within $\pm$1 subtype, as was the case between the BCD and MK classification systems. Regarding the luminosity class, we must note that the Str\"omgren photometry calibrations discriminate only between classes V and III, while the BCD procedures allow a more precise separation of classes V, IV and III.\par

\begin{figure}
   \centering
   \includegraphics[width=\hsize]{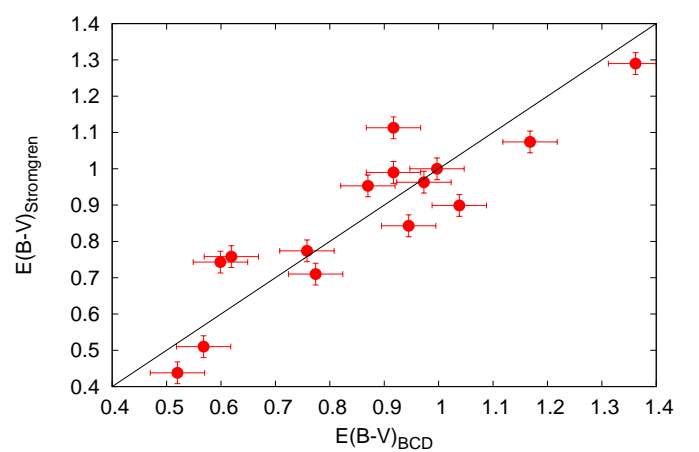}
      \caption{Comparison between the colour excesses obtained from the FAST spectra with the procedure described in this work and those obtained with 
      Str\"omgren photometry techniques. }
    \label{Stromgren}
   \end{figure}

In Fig.~\ref{Stromgren} we compare the values of $E(B-V)$ obtained with the BCD and Str\"omgren photometry techniques. For our determination we consider an error of 0.05 mag. as in the previous subsection. For the $uvby\beta$ calibration we use the error of 0.03 mag. quoted in \citet{Fabregat1998}.   The agreement is very good, with a mean difference of =0.00$\pm$0.10 mag. In both the $uvby\beta$ calibrations and the BCD system the intrinsic colours of the stars are determined mainly from the depth of the Balmer discontinuity, measured through the $c_1$ index in the $uvby$ photometric system and through the $D$ parameter in the BCD system. Both determinations are consistent and lead to similar results.\par

\begin{table*}
\caption{Comparison between the spectral classification and physical parameters obtained for 15 CBe stars and those obtained from Str{\"o}mgren-Crawford photometry calibrations. }             
\label{table:4}      
\centering          
\begin{tabular}{c c c c c c c r r  }     
\hline\hline       
             NAME        & $D(dex)$   &$ \lambda_1 (\AA)$  &  Sp/LC$_{\rm BCD}$ & Sp/LC$_{uvby\beta}$ & $E(B-V)_{\rm BCD}$ &  $E(B-V)_{uvby\beta} $  &$M_{V,BCD}  $ & $M_{V,uvby\beta} $\\
\hline 
    $ J014602.11+611502.2 $  &   0.191  &    67.2    &    B4V  &   B5-6V     &   0.917   &     1.113     &   $  -0.3 $  &$ -0.5  $   \\
    $ J014624.42+611037.3 $  &   0.225  &    68.2    &    B5V  &   B5III     &   0.758   &     0.774     &    $ -0.7  $ & $-2.3  $   \\
    $ J053237.14+260107.3 $  &    0.290  &    47.9   &    B5-6V & B4V     &   0.917    &     0.990        & $  -1.4 $ & $  -1.6    $    \\
    $ J053513.10+295912.4 $  &   0.246  &    31.8    &    B3-4III  &   B3III     &   1.168   &     1.074      &$     -2.7 $  & $ -2.5   $   \\
    $ J053554.13+295756.4 $  &   0.221  &    77.9    &    B5V  &   B5V     &   1.362   &     1.290      &   $  -1.4  $      &  $-0.8   $   \\
    $ J053654.85+301757.8 $  &   0.273  &    66.1    &    B5-6V  &   B5V     &   0.870   &     0.953      &  $   -0.2 $      &  $-0.4 $     \\
    $ J054033.87+274552.4 $  &   0.398  &    65.4    &    B8V  &   B8V     &   0.619   &     0.758      &     0.5       &  0.2      \\
    $ J054115.07+274803.2 $  &   0.139  &    68.4    &    B2V  &   B3V     &   0.997   &     1.000      &   $  -1.7  $     & $ -0.7    $  \\
    $ J054159.24+274038.1 $  &   0.254  &    74.6    &    B5-6V  &   B5V     &   0.774   &     0.710      & $    -0.4 $    &$  -0.4   $   \\
    $ J054200.54+304956.6 $  &   0.355  &    44.4    &    B7IV  &   B7V     &   0.599   &     0.743     &   $  -0.0  $  & $ -0.1     $\\
    $ J054450.31+290754.5 $  &   0.319  &    68.7    &    B7V  &   B7V     &   0.568   &     0.510     &     0.2    &  0.0     \\
    $ J054603.64+272729.5 $  &   0.211  &    48.9    &    B3IV  &   B2.5V     &   0.973   &     0.963     &  $   -1.1 $   & $-1.9 $    \\
    $ J054837.64+281710.3 $  &   0.179  &    63.4    &    B3V  &   B2V     &   0.945   &     0.843     &    $ -1.4  $  & $-0.9    $ \\
    $ J054848.26+283547.8 $  &   0.159  &    60.7    &    B2V &   B1.5-2V     &   1.038   &     0.899     &$     -1.5 $  & $-2.2 $    \\
    $ J054937.88+281123.3 $  &   0.329  &    70.3    &    B7V  &   B8V     &   0.520   &     0.438     &     0.2   & 0.2     \\

\hline                  
\end{tabular}
\end{table*}

\begin{figure}
   \centering
   \includegraphics[width=\hsize]{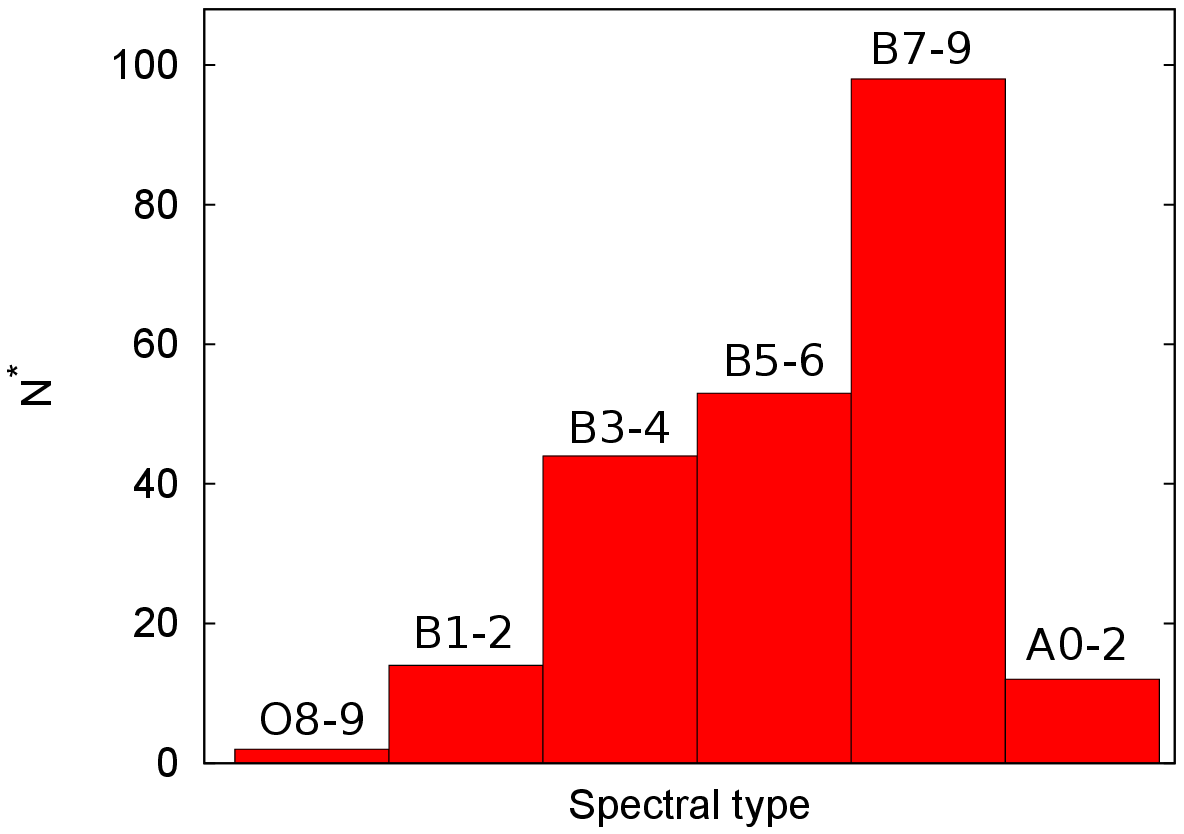}
      \caption{Histogram distribution of the spectral types.  }
         \label{Fig:SpHistogram}
   \end{figure}

\section{Analysis of the Perseus Arm area.}

As a further evaluation of our procedure, in this section we present the analysis of a sample of IPHAS follow-up spectra obtained with the FLWO/FAST spectrograph. We have selected the stars located in the Perseus arm region ($-1^{o}<b<4^{o}$,  $120^{o}<l<140^{o}$). The population of CBe stars in this area has been investigated by \citet{Raddi2013}, who analysed the INT and NOT spectra described in Sect. 3.2, and by \citet{Raddi2014} also using a sample of FAST spectra. In both papers the spectral classification was performed by comparison with standard templates, complemented with the application of standard procedures  of the MK classification system. Both works stress the difficulty of assigning luminosity classes, due to the fact that the profiles of the Balmer lines, which provide the main luminosity indicators in the considered spectral range, are contaminated by the circumstellar emission characteristic of CBe stars.\par

In the BCD system, luminosities are derived from the $\lambda_1$ parameter, which is not affected by emission. As distance determination with the spectroscopic parallax technique heavily relies on the luminosity class of the objects, it is worthwhile to reproduce the above studies with reddenings and distances computed with the techniques described in this work, in order to compare the results.\par 

Our analysis was done with 257 FAST spectra of 224 stars, that meet the criteria presented in Sect. 3.1. From each spectrum we determined the BCD ($D, \lambda_1$) parameters, the effective temperature, spectral type and luminosity class, interstellar and circumstellar colour excesses, absolute magnitude and distance. Results are presented in Tables~\ref{table:results1}  and \ref{table:results2}. It took on average 2-3 minutes of work to analyse each star by means of our semi-automatic procedure.\par 

We estimate the internal relative error in the determination of the effective temperature and distance as the difference between the individual values obtained from different spectra of the same star divided by the mean value. The mean errors measured in this way amount to 8\% in  $T_{\rm eff}$ and 24\% in distance.\par

Eighteen stars, representing 8\% of the sample, display the H$\alpha$ line in absorption. Since all the observed stars have been selected from a list of photometrically detected emission line stars \citep{witham2008}, this figure can be considered as a measure of the false detection percentage of the photometric method. It should be noted, however, that the Be phenomenon is variable, and for some stars the lack of H$\alpha$ emission may be due to a phase transition from emission to absorption in the interval between the acquisition of the photometric and spectroscopic data. The above false detection figure must be considered as an upper limit.   
 
Regarding the spectral classification, we found all sub-types from O8$-$9 up to A2. We found one Oe star, \#66, with type 
O8$-$9IVe. The presence of visible \ion{He}{II} lines in its spectrum confirms this classification. In Fig.~\ref{Fig:SpHistogram} we present the histogram of the CBe star sample distribution as a function of the spectral types.\par

\begin{figure*}
   \centering
 \includegraphics[width=18.6cm]{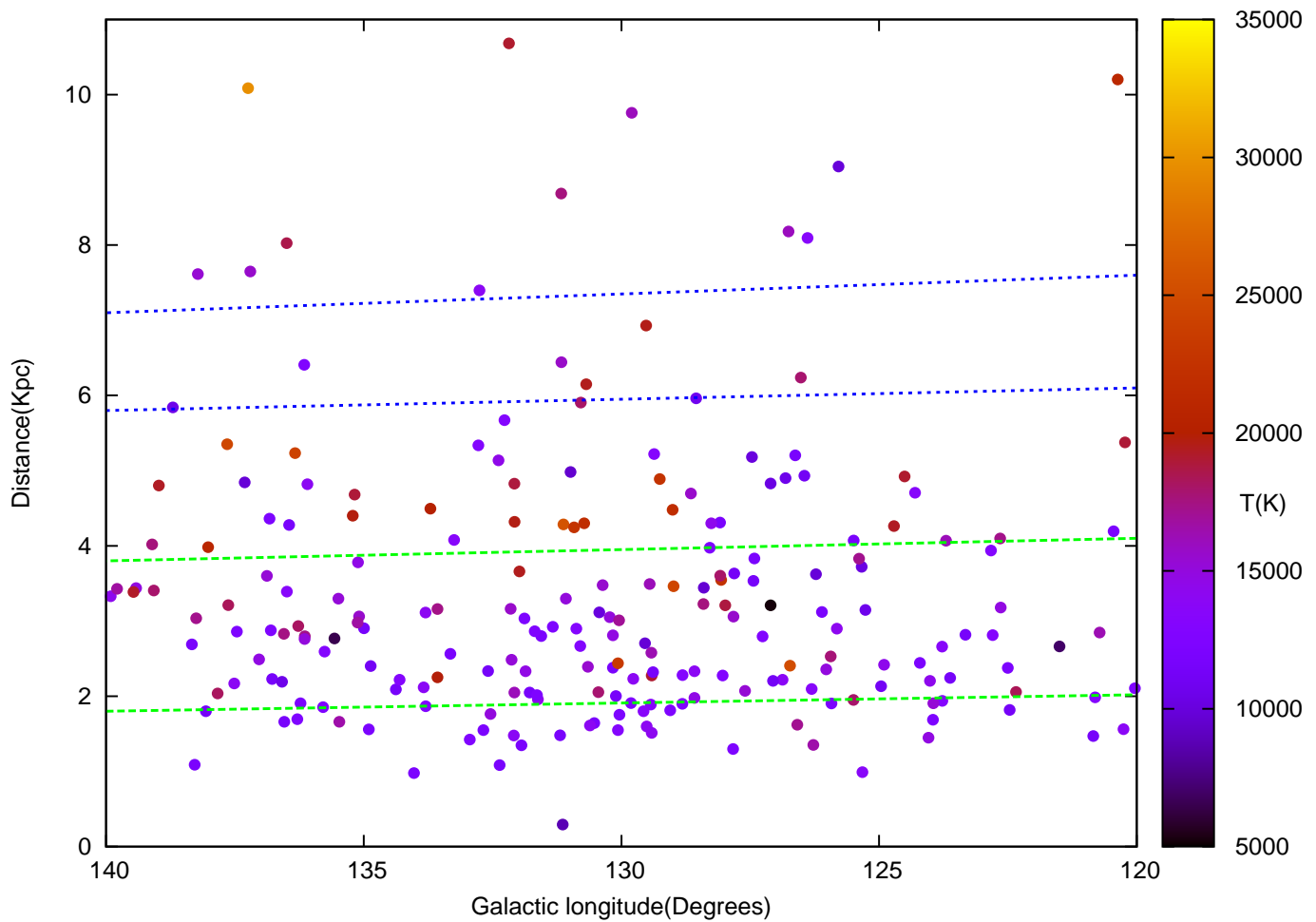}
   \caption{The distribution of the CBe stars distances as a function of the Galactic longitude. The  green and blue dashed lines mark the expected positions of the Perseus and Outer Galactic arms, respectively. Five stars are at distances too long to appear in the diagram. The colour bar at the right-hand side represents the effective temperature of the stars. }
              \label{Fig:perseus}%
    \end{figure*}

\begin{figure}
   \centering
   \includegraphics[width=\hsize]{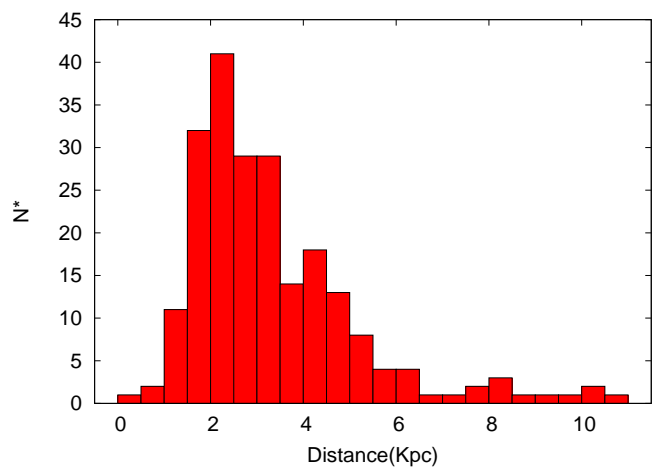}
      \caption{ Histogram distribution of the distances. }
         \label{Fig:Ravdograma}
   \end{figure}

The spatial distribution found for the CBe star sample is presented in Fig.~\ref{Fig:perseus}, where the 224 stars with distances listed in Table~\ref{table:results2} are plotted. For stars with more than one spectrum we used the mean value of the distance determinations. The locations of the Perseus and Outer Galactic arms are marked with dotted lines, following the range of distances given by \citet{Russeil2007}. Green represents the Perseus Arm at $\sim2-3.5$ kpc. and blue the Outer Arm at $\sim6-7$ kpc. The distribution of the stars does not present an apparent clustering 
in or around these two structures. Instead, they appear scattered along the two arms and the space in between, with some stars spread along larger distances, beyond the expected location of the Outer Arm. In Fig.~\ref{Fig:Ravdograma} we present an histogram of the measured distances.These results are consistent with the findings of \citet{Raddi2013}.\par

For ten stars we found distances in excess of 9 kpc., well beyond the expected location of the Outer Galactic Arm. Five of them have distances larger than 11 kpc and hence they are out of the scale of Fig.~\ref{Fig:perseus}. These distances  in some cases can be due to large errors in the absolute magnitude or the spectral classification of the stars. A detailed discussion on the uncertainties and possible biasses in distance determination based on absolute magnitudes was given in  \citet{Raddi2013}. We are, however, studying in detail, for a future work a few stars in which the large distance seems to be consistently derived from high signal to noise spectra, in order to use them as tracers of the stellar population in the outskirts of the galactic plane.\par

\section{Conclusions}

This work is the first part of a larger project to study the whole population of CBe stars photometrically detected by the IPHAS survey, and use them as galactic structure tracers. The study will be based on the analysis of follow-up spectroscopy obtained at the FLWO telescope with the FAST spectrograph.\par

In this paper we have presented the method devised to obtain the spectral classification and the astrophysical parameters of the stars from the FLWO/FAST spectra, using the  semi-automatic procedure we developed  based on the BCD (Barbier-Chalonge-Divan) spectrophotometric system.\par

We have validated the method by comparing its results for two samples of CBe stars with independent results for the same stars obtained with different astrophysical techniques. In particular, we compared our results with those obtained with spectral template fitting and MK classification system standard techniques for one of the samples, and with 
Str\"omgren photometry standard calibrations for the other. In both cases we obtained a general good agreement in the spectral classification, within two spectral subtypes and one luminosity class for most of the stars. We have also compared our results on the interstellar extinction with those of the two samples referred to above, obtaining a good agreement in both cases. \par

We have also analysed a sample of CBe stars in the direction of the Perseus Arm  ($-1^{o}<b<4^{o}$,  $120^{o}<l<140^{o}$).  We didn't find any significant clustering of stars at the expected distances of the Perseus and Outer Arms. Even with over three times the number of stars considered by \citet{Raddi2013} we obtain the same negative result, indicating that the errors involved in the spectroscopic parallax and absolute magnitude determination blur any hint of Galactic spiral structure, if indeed the CBe stars trace it.\par

\begin{acknowledgements}
This paper made use of spectroscopic data that were obtained at the FLWO-1.5m with FAST, which is operated by Harvard-Smithsonian Centre for Astrophysics. In particular we want to thank Perry Berlind and Mike Calkins for their role in obtaining most of the FLWO-1.5m/FAST data.\par

The paper also makes use of data obtained as part of the INT Photometric H$\alpha$ Survey of the Northern Galactic Plane (IPHAS) carried out at the Isaac Newton Telescope (INT). The INT is operated on the island of La Palma by the Isaac Newton Group in the Observatorio del Roque de los Muchachos of the Instituto de Astrof\'\i isica de Canarias. All IPHAS data are processed by the Cambridge Astronomical Survey Unit, at the Institute of Astronomy in Cambridge. We also acknowledge the use of data obtained at the INT and the Nordic Optical Telescope as part of a CCI International Time Programme.\par

The work of L.G. and J.F. is supported by the Spanish Ministerio de Econom\'\i a y Competitividad, and FEDER, under contracts AYA2010-18352 and AYA2013-48623-C2-2-P, and by the Generalitat Valenciana project of excellence PROMETEOII/2014/060. R.R. received funding from the European Research Council under the European Union's Seventh Framework Programme (FP/2007-2013)/ERC Grant Agreement no. 320964 (WDTracer).  N.J.W. was supported by a Royal Astronomical Society fellowship. Finally, we are grateful to  Dr. Pablo Reig for his assistance in the classification  in the MK system. 
\end{acknowledgements}


\begin{table*}[b]
\caption{Photometric $r$ magnitudes, H$\alpha$EW, BCD parameters, effective temperatures and spectral classification for the stars studied in this work. The parameters of stars with multiple spectra have been measured from each individual spectrum. They appear as multiple entries, with the same correlative number in the first column.}       
\label{table:results1}      
\centering          
\begin{tabular}{r c c r r r r r r c }     
\hline\hline       
  \#        & IPHAS NAME  &$l$(deg)  & $b$(deg) &   $r$(mag)& H$\alpha$EW(\AA) & $D$(dex)      &  $\lambda_1(\AA) $& $T_{\rm eff}(K)$ & SpT/LC  \\
\hline 

1	&	J002540.04+623203.2	&	119.9	&	$-$0.18	&	14.46	&	$-$11.18	&	0.318	&	58.4	&	13186	&	B7$-$8V	\\
2	&	J002754.61+603001.4	&	120.0	&	$-$2.23	&	13.66	&	$-$11.52	&	0.345	&	56.9	&	12380	&	B7$-$8 V    	\\
3	&	J002758.97+622906.1	&	120.2	&	$-$0.26	&	15.03	&	$-$16.39	&	0.206	&	59.4	&	18966	&	B3$-$4V	\\
4	&	J002843.24+615216.2	&	120.2	&	$-$0.88	&	14.38	&	$-$14.33	&	0.294	&	65.5	&	13778	&	B7 V	\\
5	&	J003000.79+612238.8	&	120.3	&	$-$1.38	&	14.24	&	$-$11.72	&	0.155	&	44.6	&	21195	&	B2 III    	\\
6	&	J003025.04+645500.0	&	120.7	&	2.13	&	15.95	&	$-$28.07	&	0.086	&	52.6	&	29473	&	B1 IV	\\
7	&	J003055.94+610048.8	&	120.4	&	$-$1.75	&	15.39	&	$-$9.64	&	0.376	&	49.9	&	11980	&	B7$-$8 IV   	\\
\hline                  
\end{tabular}
\end{table*}

\begin{table*}[b]
\caption*{Table 5. Continued... }              

\centering          
\begin{tabular}{r c c r r r r r r c }     
\hline\hline       
  \#        & IPHAS NAME  &$l$(deg)  & $b$(deg) &   $r$(mag)& H$\alpha$EW(\AA) & $D$(dex)      &  $\lambda_1(\AA) $& $T_{\rm eff}(K)$ & SpT/LC  \\
\hline

59	&	J012536.96+593053.0	&	127.2	&	$-$3.06	&	13.74	&	7.81	&	0.303	&	70.1	&	13123	&	B7 V	\\
60	&	J012540.57+623025.7	&	126.8	&	$-$0.10	&	13.42	&	$-$24.40	&	0.278	&	79.4	&	13801	&	B5$-$6 V    	\\
61	&	J012611.30+585127.4	&	127.4	&	$-$3.71	&	15.54	&	9.05	&	0.355	&	74.5	&	11820	&	B7$-$8 V  	\\
62	&	J012634.69+641850.9	&	126.7	&	1.70	&	12.77	&	$-$17.65	&	0.136	&	69.0	&	25369	&	B1 V	\\
63	&	J012657.63+591436.3	&	127.4	&	$-$3.31	&	15.52	&	5.21	&	0.450	&	70.3	&	10251	&	A0$-$1 V  	\\
64	&	J012745.08+625154.3	&	127.0	&	0.28	&	13.46	&	$-$18.13	&	0.403	&	69.4	&	11182	&	B9 V	\\
65	&	J012821.12+635754.0	&	126.9	&	1.38	&	14.63	&	1.88	&	0.069	&	62.2	&	33482	&	O8$-$9 IV   	\\
\hline                  
\end{tabular}
\end{table*}

\begin{table*}[b]
\caption*{Table 5. Continued... }              

\centering          
\begin{tabular}{r c c r r r r r r c }     
\hline\hline       
  \#        & IPHAS NAME  &$l$(deg)  & $b$(deg) &   $r$(mag)& H$\alpha$EW(\AA) & $D$(dex)      &  $\lambda_1(\AA) $& $T_{\rm eff}(K)$ & SpT/LC  \\
\hline

116	&	J015939.01+643615.4	&	130.1	&	2.69	&	13.26	&	$-$18.10	&	0.282	&	50.5	&	14790	&	B5$-$6 V    	\\
116	&	J015939.01+643615.4	&	130.1	&	2.69	&	13.26	&	$-$21.62	&	0.279	&	61.6	&	14550	&	B5$-$6 V    	\\
117	&	J020037.84+652133.9	&	130.0	&	3.45	&	13.08	&	3.84	&	0.151	&	71.9	&	23301	&	B2 V	\\
118	&	J020048.75+585835.1	&	131.7	&	$-$2.69	&	13.58	&	$-$13.89	&	0.375	&	68.3	&	11640	&	B9 V	\\
118	&	J020048.75+585835.1	&	131.7	&	$-$2.69	&	13.58	&	$-$14.27	&	0.335	&	72.4	&	12174	&	B8 V	\\
119	&	J020049.45+635943.9	&	130.4	&	2.14	&	13.91	&	$-$28.90	&	0.223	&	58.6	&	17824	&	B3$-$4V	\\
120	&	J020056.02+575529.3	&	132.0	&	$-$3.70	&	14.99	&	$-$9.65	&	0.204	&	74.9	&	18827	&	B4 V	\\
\hline                  
\end{tabular}
\end{table*}

\begin{table*}[b]
\caption*{Table 5. Continued... }              

\centering          
\begin{tabular}{r c c r r r r r r c }     
\hline\hline       
  \#        & IPHAS NAME  &$l$(deg)  & $b$(deg) &   $r$(mag)& H$\alpha$EW(\AA) & $D$(dex)      &  $\lambda_1(\AA) $& $T_{\rm eff}(K)$ & SpT/LC  \\
\hline

167	&	J023235.12+640522.8	&	133.7	&	3.35	&	13.52	&	$-$37.29	&	0.180	&	46.8	&	20188	&	B3$-$4 IV   	\\
168	&	J023410.30+612440.6	&	134.9	&	0.95	&	13.64	&	$-$92.45	&	0.336	&	80.0	&	11985	&	B8 V	\\
169	&	J023411.99+595634.4	&	135.4	&	$-$0.40	&	13.13	&	$-$17.33	&	0.241	&	74.5	&	16334	&	B5$-$6 V    	\\
169	&	J023411.99+595634.4	&	135.4	&	$-$0.40	&	13.13	&	$-$16.63	&	0.241	&	69.0	&	16524	&	B5 V	\\
170	&	J023758.12+634635.6	&	134.3	&	3.29	&	13.30	&	$-$8.86	&	0.388	&	63.0	&	11554	&	B9 V	\\
170	&	J023758.12+634635.6	&	134.3	&	3.29	&	13.30	&	$-$9.42	&	0.435	&	55.5	&	10930	&	B9 V	\\
171	&	J023809.91+620224.6	&	135.0	&	1.71	&	13.20	&	$-$14.26	&	0.244	&	51.6	&	16531	&	B3$-$4 IV   	\\
\hline                  
\end{tabular}
\end{table*}

\begin{table*}[b]
\caption*{Table 5. Continued... }              

\centering          
\begin{tabular}{r c c r r r r r r c }     
\hline\hline       
  \#        & IPHAS NAME  &$l$(deg)  & $b$(deg) &   $r$(mag)& H$\alpha$EW(\AA) & $D$(dex)      &  $\lambda_1(\AA) $& $T_{\rm eff}(K)$ & SpT/LC  \\
\hline

213	&	J030124.02+610221.3	&	138.0	&	2.03	&	14.85	&	$-$29.58	&	0.174	&	73.2	&	21026	&	B2 V	\\
214	&	J030144.21+632853.5	&	136.8	&	4.19	&	14.39	&	$-$7.81	&	0.268	&	57.1	&	15258	&	B5$-$6 V    	\\
214	&	J030144.21+632853.5	&	136.8	&	4.19	&	14.39	&	$-$7.89	&	0.272	&	61.8	&	14835	&	B5$-$6 V    	\\
215	&	J030317.45+583402.0	&	139.4	&	$-$0.01	&	14.57	&	$-$20.15	&	0.280	&	44.0	&	14877	&	B5$-$6 IV   	\\
216	&	J030331.33+585234.6	&	139.2	&	0.26	&	14.64	&	$-$8.18	&	0.111	&	35.5	&	20644	&	B2 Ib	\\
217	&	J030332.31+623856.7	&	137.4	&	3.56	&	14.32	&	$-$8.93	&	0.356	&	50.9	&	12303	&	B7$-$8 V    	\\
217	&	J030332.31+623856.7	&	137.4	&	3.56	&	14.32	&	$-$8.98	&	0.393	&	59.1	&	11562	&	B9 V	\\
\hline                  
\end{tabular}
\end{table*}
%


\begin{table*}[b]
\caption{Spectroscopic gradient, colour excess, absolute and intrinsic magnitudes and distances for the stars studied in this work.}          
\label{table:results2}      
\centering          
\begin{tabular}{r c c r r c c c c c c }     
\hline\hline       
        \#        & $\Phi_b$ &$ \Phi_b^0$  &  $M_V$  & $E(B-V) $ &   $A_V$& $A_r$ & $f_D$      &  $E_{cs}(B-V) $ & $\Delta r$ & $d$(kpc)  \\
\hline 

1	&	2.216	 &	 0.810	 &	 0.00	 &	 0.952	 &	 2.953	  &	  2.480  &	 0.03  &       0.022&	    0.084   &	    	    2.53  	\\
2	&	2.009	 &	 0.825   &	 0.00	 &	 0.802	 &	 2.486	  &	  2.088  &	 0.03  &       0.023&	    0.085   &	    	    2.10 	\\
3	&	2.430	 &	 0.740	 &	 $-$1.50	 &	 1.145	 &	 3.550	  &	  2.982  &	 0.05  &       0.032&	    0.081   &	   	    5.37 	\\
4	&	2.760	 &	 0.800	 &	 0.00	 &	 1.328	 &	 4.117	  &	  3.458  &	 0.04  &       0.028&	    0.084   &	    	    1.56  	\\
5	&	2.207	 &	 0.720	 &	 $-$3.25   &	 1.007	 &	 3.123	  &	  2.623  &	 0.03  &       0.023&	    0.080   &	    	   10.20	\\
6	&	2.251	 &	 0.675   &	 $-$4.75   &	 1.067	 &	 3.309	  &	  2.780  &	 0.09  &       0.056&	    0.152   &	    	   44.30	\\
7	&	2.457	 &	 0.845   &	 $-$0.50	 &	 1.092	 &	 3.385	  &	  2.843  &	 0.03  &       0.019&	    0.085   &	    	    4.19  	\\
\hline                  
\end{tabular}
\end{table*}

\begin{table*}[b]
\caption*{Table 6. Continued... }             
      
\centering          
\begin{tabular}{r c c r r c c c c c c }     
\hline\hline       
        \#        & $\Phi_b$ &$ \Phi_b^0$  &  $M_V$  & $E(B-V) $ &   $A_V$& $A_r$ & $f_D$      &  $E_{cs}(B-V) $ & $\Delta r$ & $d$(kpc)  \\
\hline 

59	&	1.343	 &	 0.806	 &	 0.50	 &	 0.363	 &	 1.127	  &	  0.947  &	0.00 &       0.000&	    0.000  &	   	    2.79      \\
60	&	2.078	 &	 0.898   &	 $-$0.25   &	 0.799	 &	 2.478	  &	  2.081  &	 0.08  &       0.048&	    0.171   &	    	    2.21 	\\
61	&	1.880	 &	 0.860	 &	 0.75	 &	 0.691	 &	 2.143	  &	  1.800  &	0.00 &       0.000&	    0.000   &	    	    3.83  	\\
62	&	2.442	 &	 0.699	 &	 $-$2.00	 &	 1.180	 &	 3.659	  &	  3.074  &	 0.05  &       0.035&	    0.165   &	    	    2.40  	\\
63	&	1.568	 &	 0.930	 &	 0.75	 &	 0.432	 &	 1.340	  &	  1.125  &	0.00 &       0.000&	    0.000  &	   	    5.18       \\
64	&	1.724	 &	 0.895   &	 0.75	 &	 0.562	 &	 1.742	  &	  1.463  &	 0.06  &       0.036&	    0.122   &	    	    1.82 	\\
65	&	2.323	 &	 0.670	 &	 $-$4.50	 &	 1.120	 &	 3.472	  &	  2.916  &	0.00 &       0.000&	    0.000  &	   	   18.73      \\
\hline                  
\end{tabular}
\end{table*}

\begin{table*}[b]
\caption*{Table 6. Continued... }             
      
\centering          
\begin{tabular}{r c c r r c c c c c c }     
\hline\hline       
        \#        & $\Phi_b$ &$ \Phi_b^0$  &  $M_V$  & $E(B-V) $ &   $A_V$& $A_r$ & $f_D$      &  $E_{cs}(B-V) $ & $\Delta r$ & $d$(kpc)  \\
\hline 

116	&	1.995	 &	 0.785   &	 $-$1.00	 &	 0.820	 &	 2.542	  &	  2.136  &	 0.06  &       0.036&	    0.083   &	    	    2.76 	\\
116	&	1.655	 &	 0.785   &	 $-$0.50	 &	 0.589	 &	 1.828	  &	  1.536  &	 0.07  &       0.043&	    0.083   &	    	    2.86 	\\
117	&	2.370	 &	 0.710	 &	 $-$1.75   &	 1.124	 &	 3.487	  &	  2.929  &	0.00 &       0.000&	    0.000  &	   	    2.43      \\
118	&	1.603	 &	 0.865   &	 0.75	 &	 0.500	 &	 1.551	  &	  1.303  &	 0.04  &       0.027&	    0.086   &	    	    2.03 	\\
118	&	1.656	 &	 0.845   &	 0.75	 &	 0.549	 &	 1.704	  &	  1.432  &	 0.04  &       0.028&	    0.085   &	    	    1.91 	\\
119	&	2.733	 &	 0.740	 &	 $-$1.00	 &	 1.350	 &	 4.185	  &	  3.515  &	 0.09  &       0.057&	    0.166   &	    	    2.05  	\\
120	&	2.250	 &	 0.745   &	 $-$1.00	 &	 1.019	 &	 3.161	  &	  2.655  &	 0.03  &       0.019&	    0.081   &	    	    4.82 	\\
\hline                  
\end{tabular}
\end{table*}

\begin{table*}[b]
\caption*{Table 6. Continued... }             
      
\centering          
\begin{tabular}{r c c r r c c c c c c }     
\hline\hline       
        \#        & $\Phi_b$ &$ \Phi_b^0$  &  $M_V$  & $E(B-V) $ &   $A_V$& $A_r$ & $f_D$      &  $E_{cs}(B-V) $ & $\Delta r$ & $d$(kpc)  \\
\hline 

167	&	2.422	 &	 0.730	 &	 $-$2.50	 &	 1.146	 &	 3.553	  &	  2.985  &	 0.12  &       0.074&	    0.163   &	    	    4.49 	\\
168	&	2.363	 &	 0.860	 &	 0.50	 &	 1.018	 &	 3.156	  &	  2.651  &	 0.30  &       0.184&	    0.537   &	    	    1.55 	\\
169	&	2.376	 &	 0.765   &	 $-$0.50	 &	 1.091	 &	 3.385	  &	  2.843  &	 0.05  &       0.034&	    0.082   &	    	    1.47  	\\
169	&	2.238	 &	 0.760	 &	 $-$0.75   &	 1.001	 &	 3.105	  &	  2.608  &	 0.05  &       0.033&	    0.082   &	    	    1.85  	\\
170	&	1.543	 &	 0.865   &	 0.75	 &	 0.459	 &	 1.425	  &	  1.197  &	 0.02  &       0.017&	    0.086   &	    	    1.87  	\\
170	&	1.658	 &	 0.890	 &	 0.15	 &	 0.520	 &	 1.613	  &	  1.354  &	 0.03  &       0.018&	    0.086   &	    	    2.32  	\\
171	&	1.979	 &	 0.765   &	 $-$1.50	 &	 0.822	 &	 2.550	  &	  2.142  &	 0.04  &       0.028&	    0.082   &	    	    3.40 	\\
\hline                  
\end{tabular}
\end{table*}

\begin{table*}[b]
\caption*{Table 6. Continued... }             
      
\centering          
\begin{tabular}{r c c r r c c c c c c }     
\hline\hline       
        \#        & $\Phi_b$ &$ \Phi_b^0$  &  $M_V$  & $E(B-V) $ &   $A_V$& $A_r$ & $f_D$      &  $E_{cs}(B-V) $ & $\Delta r$ & $d$(kpc)  \\
\hline 

213	&	2.581	 &	 0.725   &	 $-$1.25   &	 1.257	 &	 3.898	  &	  3.274  &	 0.09  &       0.059&	    0.162   &	    	    3.98 	\\
214	&	1.939	 &	 0.780	 &	 $-$0.50	 &	 0.785	 &	 2.434	  &	  2.045  &	 0.02  &       0.015&	    0.083   &	    	    3.81  	\\
214	&	2.087	 &	 0.784   &	 $-$0.50	 &	 0.883	 &	 2.737	  &	  2.299  &	 0.02  &       0.015 &	    0.083   &	    	    3.39 	\\
215	&	2.751	 &	 0.770	 &	 $-$1.50   &	 1.342	 &	 4.161	  &	  3.496  &	 0.06  &       0.040&	    0.083   &	    	    3.43  	\\
216	&	2.752	 &	 0.704	 &	 $-$5.50	 &	 1.386	 &	 4.299	  &	  3.611  &	 0.02  &       0.016&	    0.000   &	    	   22.08 	\\
217	&	2.033	 &	 0.835   &	 $-$0.50	 &	 0.811	 &	 2.515	  &	  2.113  &	 0.02  &       0.017&	    0.085   &	    	    3.58  	\\
217	&	2.250	 &	 0.860	 &	 0.25	 &	 0.941	 &	 2.919	  &	  2.452  &	 0.02  &       0.017&	    0.086   &	    	    2.14 	\\
\hline                  
\end{tabular}
\end{table*}

\bibliographystyle{aa} 
\bibliography{refs} 


\appendix

\section{The determination of the BCD parameters in CBe stars}

After the discovery of the two-component Balmer discontinuity in $\zeta$~Tau by \citet{ch39}, the long series of observations carried on classical Be stars in the BCD system revealed that this feature is a common phenomenon among CBe stars. The so-called first, or photospheric-like component of the BD, resembles that of a normal B-type star, whilst the second one is in emission or in absorption relative to the normal level of the
Balmer continuum, depending on the Be-phase of the object. The strongest absorption or emission in this second
component of the BD always produces at $\lambda\simeq3647\ \AA$, close to the theoretical limit of the Balmer-line series, which indicates that it is formed in an environment where the gas pressure is significantly lower than in the stellar photosphere.\par 

  Except for $\gamma$~Cas (O9Ve, HD 5394), during its huge emission episodes or outbursts, in 1935 and 1938 \citep{bb48}, in all other CBe stars observed more or less regularity, no changes have been detected in the value of the first component of the BD ($D_*$) within the limits of uncertainties that characterize the BCD system (0.015 dex in $D$ and 1-2 {\AA }  in $\lambda_1$). This has been proved, in particular, in some iconic CBe stars in the northern hemisphere as X~Per (O9.5Ve, HD 24534), Pleione (B8IV-Ve, HD 23862), $\zeta$~Tau (B2IIIe, HD 37202), 48~Lib (B3-4IVe, HD 142983), 88~Her (B6IVe, HD 162732), $\chi$~Oph (B0Ve, HD 148184), and 59~Cyg (B1Ve, HD 200120) \citep{loore79,zor83,zor86,zor89,zor82a,zor82b,Zorec1991}, but also in some frequently observed CBe stars in the southern hemisphere, in par\-ti\-cu\-lar $\alpha$~Eri (B5III3, HD 10144) \citep{vin06,coch14}. The constancy of $D_*$ has also been proved in stars that underwent Be$\rightleftarrows$B$\rightleftarrows$Be-shell-phase changes such as $\gamma$~Cas, Pleione, 88~Her and 59~Cyg.\par 
  
  Calling total BD the quantity $D=D_*+\delta D$ (dex), the spectrophotometric variations of Be stars are currently different according to the CBe phases. Increasing emissions phases ($\delta D<0$) are generally accompanied by a brightening and a reddening of the Paschen continuum, while shell phases ($\delta D>0$) are generally characterized  by little decreases of the stellar brightening, if any, but almost no change in the colour of the Paschen continuum. An extensive compilation of these spectrophotometric behaviours was published by \citet{mouj98b}.\par
  
  As noted earlier in this appendix, the photospheric component of the BD defined in the BCD system corresponds to the ratio between an upper flux, $F^+_{\lambda_D}$, and a lower radiation flux, $F^-_{\lambda_D}$ determined at $\lambda_{\rm D}\simeq3700\ \AA$. It is worth insisting on some characteristics of these two fluxes when dealing with CBe stars: \par
  
  1) $F^+_{\lambda_D}$ is an extrapolated flux of the Paschen continuum, whose purpose is to represent the energy distribution that would exist if the wings of the Stark-broadened Balmer lines $Hn$ with $n>8$ would not overlap each other to end up producing a pseudo-continuum energy distribution in low resolution spectra well before $\lambda_{\rm D}=3648\ \AA$; \par
  
 2) The pseudo-continuum thus formed rejoins the level of the Balmer continuum just shortly after the BD in normal O-, B-, A-, F-type stars, that is, objects without extra emission or absorption produced by circumstellar gaseous environments; \par
 
 3) $F^-_{\lambda_D}$ belongs to the aforementioned pseudo-continuum, which thus also lies on the Paschen continuum. It does mean that both fluxes $F^{\pm}_{\lambda_D}$ are affected by the same amount of emission or absorption, 
$\Delta F_{\lambda_D}$, raised in the circumstellar envelope (or disc) by the electron scattering, bound-free and free-free transitions mostly of hydrogen and helium atoms. This is the main reason that justifies our care to identify the point were the last lines of the Balmer series overlap. In this way, the circumstellar emission or absorption perturbing the Balmer side of the stellar continuum has no incidence on the the determination of the photospheric BD, $D_*$.\par
   The $F^-_{\lambda_D}$ can be affected by the emission in the highest members of the Balmer line series, but
this phenomenon occurs rarely; it happened perhaps in $\gamma$~Cas during its historic outbursts that we mentioned earlier.\par
   However, the empirical determination of $D_*$ cannot avoid two effects related with the presence of the 
circumstellar environment:\par

   a) The genuine photosheric fluxes $F^{\pm}_{\lambda_D}$ transform into $F^{\pm}_{\lambda_D}+\Delta F_{\lambda_D}$, which implies that:

\begin{equation}
D^{\rm obs}_* \simeq D^o_*+\left(\frac{\Delta F_{\lambda_D}}{F^+_{\lambda_D}}\right)(1-10^{D^o_*}),
\label{eq1}
\end{equation}

\noindent where $D^o_*$ is the actual photospheric BD, and according to the models we present below, in the hottest stars where these effects can be the highest, it is $\Delta F_{\lambda_D}/F_{\lambda^{+}_D}\lesssim0.05$;\par

   b) Since $F^+_{\lambda_D}$ is extrapolated, it is affected by the colour changes of the Paschen continuum due to the $\lambda^3$-dependency of the circumstellar envelope-opacity.\par
   
\begin{figure*}[ht!]
\center\includegraphics[]{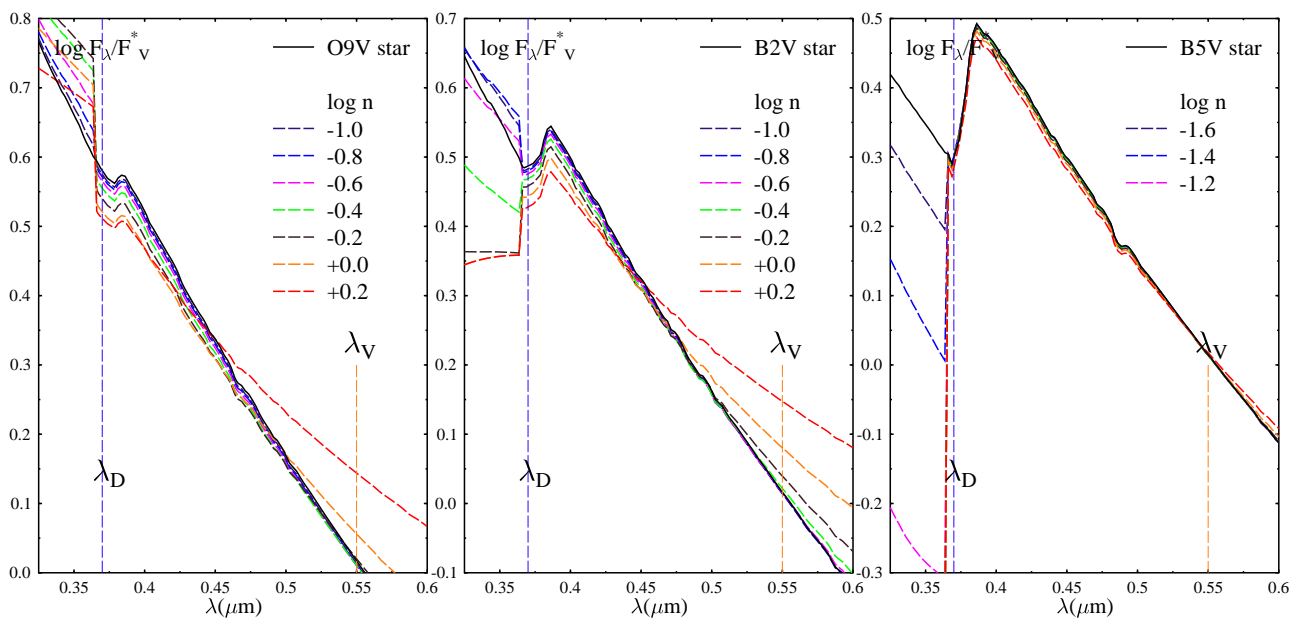}
\caption{\label{be-sh} Changes of the energy distribution in the spectral interval $0.325\leq\lambda\leq0.6$ 
$\mu$m induced by a circumstellar disc around stars of spectral type O9V, B2V and B5V. In the figure are indicated the wavelength $\lambda_D$ at which is measured $D_*$, and the central wavelength $\lambda_V$ of the $V$ magnitude. In the column $\log n$, $n$ is given in units of  $N/10^{12}$ with $[N]=$ cm$^{-3}$}
\end{figure*}

   \citet{mouj98a}, \citet{mouj99,mouj00a} and \citet{mouj00b} have sketched and interpreted the long-term spectrophotometric changes of CBe stars to estimate the physical parameters characterizing their circumstellar envelopes. Similar, but somewhat more detailed interpretations of these behaviours, were studied in $\zeta$~Tau by \citet{car09}. To have an overview of the circumstellar effects on the measured $D_*$ as well as to quantify these effects, we have produced the models shown in Fig.~\ref{be-sh}. These calculations are based on the circumstellar disc-model discussed in \citet{zor07}, which were adapted to produce non-LTE visible continua. We oriented the disc at $i=70^o$ to easily produce Be and Be-Shell phases in the visual energy distributions by changing only the disc density near the central star, which is here parametrized with the particle number-density $n=N/10^{12}$ with $[N]=$ cm$^{-3}$.\par
 
 To simplify the calculations, we have adopted the average radial temperature distribution in the circumstellar disc derived in \citet{mouj00b}, while its density follows the power-law deduced in \citet{zor07} that was needed to account for the \ion{F}{ii} emission lines. For the models shown in Fig.~\ref{be-sh} we used stellar spectra given by \citet{kur93} for: $T_{\rm eff}=30\,000$ K, $\log g=4.0$ (O9V); $T_{\rm eff}=22\,500$ K, $\log g=4.0$ (B2V); and $T_{\rm eff}=15\,000$ K, $\log g=4.0$ (B5V). In these models, the spectral region near $F^+_{\lambda_D}$ is only sketched to avoid detailed calculations of the level occupation probabilities in the upper hydrogen atomic level and the corresponding truncation of partition functions \citep{humi88} that would necessarily allow for the still badly-mastered non-ideal effects \citep{rozo06}. \par

\begin{table}[t]
\centering
\caption[]{\label{table_D} Measurement errors induced by the circumstellar emission/absorption on the
estimate of the photospheric component $D_*$ BD and on the corresponding absolute magnitude $M_{\rm V}(\lambda_1,D_*)$ determination.}
\tabcolsep 1.8pt
\scalebox{0.94}{\begin{tabular}{lr|rrrrrrr}
\hline
\multicolumn{2}{c}{$\log N/10^{12}$} & $-1.0$ & $-0.8$ & $-0.6$ & $-0.4$ & $-0.2$ & $0.0$ & $+0.2$  \\ 
 Sp.T & $D_*$ & \multicolumn{7}{c}{$D^{\rm obs}_*-D_*$} \\
O8V  & 0.049 & 0.000 & 0.000    & $-0.001$ & $-0.002$ & $-0.005$ & $-0.012$ & $-0.023$ \\ 
B2V  & 0.130 & 0.000 & $-0.001$ & $-0.002$ & $-0.004$ & $-0.009$ & $-0.020$ & $-0.038$ \\
\multicolumn{2}{c}{$\log N/10^{12}$}  & $-1.6$ & $-1.4$ & $-1.2$ & $-1.0$ & $-0.8$ & $-0.6$ & $-0.4$   \\
B5V  & 0.261 & 0.000 & 0.000    & 0.000    & 0.000    & $-0.001$ & $-0.002$ & $-0.005$ \\ 
\hline
     && \multicolumn{7}{c}{$\delta M_{\rm V}$} \\
O8V    &  & $-0.01$ & $-0.02$ & $-0.02$ & $-0.04$ & $-0.09$ & $-0.21$ & $-0.41$ \\ 
O8III  &  & $-0.01$ & $-0.01$ & $-0.00$ & $-0.03$ & $-0.07$ & $-0.16$ & $-0.30$ \\ 
B2V    &  & $-0.01$ & $-0.02$ & $-0.02$ & $-0.06$ & $-0.14$ & $-0.30$ & $-0.60$ \\
B2III  &  & $-0.01$ & $-0.02$ & $-0.02$ & $-0.04$ & $-0.10$ & $-0.23$ & $-0.45$ \\
B5V    &  & 0.00    & 0.00    & 0.00    & 0.00    & $-0.01$ & $-0.02$ & $-0.06$ \\ 
B5III  &  & 0.00    & 0.00    & 0.00    & 0.00    & $-0.01$ & $-0.02$ & $-0.05$ \\ 
\hline
\end{tabular}}
\end{table}

  The spectrophotometric behaviours shown in Fig.~\ref{be-sh} are an excerpt of many others that can be
obtained just by changing the aspect angle $i$. The amount of continuum emission and absorption also depends
on the size and opening angle of the lemniscat-like shaped circumstellar envelope. As much as it concerns the present models, it is rather difficult to obtain shell-phase in very hot stars. On the contrary, Be and shell phases can be obtained easily for B2-type objects, and late type B would rather display shell aspects. According to the effective temperature among early type CBe stars, the emission in the second BD is accompanied either by a reddening or a blueing of the Paschen continuum. In late type CBe stars, the shell-absorption seems not to be accompanied by any significant change of in colour of the Paschen continuum energy distribution. In all these cases, we can note that when the emission or the absorption in the second BD vary a lot, the phospheric energy jump at the BD remains remarkably unchanged.\par 

In Table~\ref{table_D} we present the values of $D_*$ for the unperturbed stars, as well as the errors $\delta D_*=D_*^{\rm obs}-D_*$ committed on the $D_*$-determination carried by the effects a) and b) described above. Because we are not interested in giving detailed physical interpretations of the calculated effects, the parameter $\log n$ must be regarded only as a way to produce a given amount of continuum emission or absorption. We should thus pay attention only on these amounts and conclude that the events labeled with $\log n= 0.0$ and $+0.2$ are rather extreme and rarely observed. In this respect, we note that in the compilation made by \citet{mouj98b}, which actually corresponds to spotted measurements concerning otherwise long-term Be-star "bumper" activity \citep{cook95,hb98,hb00,kell02,menn02,
dw06}, the magnitude changes are of the order of $0.1\lesssim\Delta V\lesssim0.2$ mag and rarely as high as 
$\Delta V\lesssim0.5$ mag.\par

   In the lower part of Table~\ref{table_D} we also give the errors $\delta M_{\rm V}$ that $\delta D_*$ values would produce on the visual absolute magnitude $M_{\rm V}(\lambda_1,D_*)$. We notice that in all cases the $\delta M_{\rm V}$ errors are smaller than the amplitude of $M_{\rm V}(\lambda_1,D_*)$ values that may concern a single average MK spectral type \citep{Zorec1991}.\par
   Let us finally mention that errors made on the estimate of $D_*$ can be somewhat reduced if the colour effect
acting on the extrapolated $F^+_{\lambda_D}$ flux, which resembles that of an increased interstellar extinction, is taken into account by using

\begin{eqnarray}
\begin{array}{rcl}
\displaystyle D_* & = & D_*^{\rm obs}+\Delta D_* \\
\displaystyle \Delta D_* & = & a\times[\Phi^{\rm obs}-\Phi(\lambda_1,D^o_*)],
\end{array}
\label{corr_d}
\end{eqnarray}

\noindent where $a=0.02$, if the Paschen colour gradient is $\Phi=\Phi_{\rm rb}$ of the BCD system. This relation was derived by the creators of the BCD system \citep{div54}, and since then applied to every determination of the BD when the ISM extinction detected by a high gradient difference $\Delta\Phi=\Phi^{\rm obs}-\Phi(\lambda_1,D_*)]$. In extreme cases $\Delta\Phi\simeq0.5$ $\mu$m, the use of
Eq.~(\ref{corr_d}) can reduce the largest uncertainties $\delta D_*$ given in Table~\ref{table_D} to the the standard BCD uncertainty $\delta D\lesssim0.015$ dex.\par
 It is finally worth noting that the change of $\lambda_1$ due to $\delta D_*$ and the variation of the slope of the Paschen continuum is extremely small and thus negligible.\par



\section{Evaluation of the flux calibration of the FLWO/FAST spectra}

The determination of reliable physical parameters of stars by means of the BCD system methods requires the spectra to have at least an accurate relative flux calibration.  In this appendix we present an evaluation of the flux calibration of the FLWO/FAST sample, by comparing different spectra of the same objects. In the discussion below we analyse only the spectral region between 3\,700 and 4\,600 $\AA$, where the BCD parameters are measured. For some spectra drops in the signal-to-noise and large differences in the flux calibrations at shorter wavelengths are apparent. However, as explained in Sects. 2.1 and 3.1, they don't have any impact in the determination of the physical parameters of the stars.

\begin{figure*}[b]
   \centering
   \includegraphics[angle=90, trim={1.1cm 0.0cm 0.1cm -1.5cm},clip,width=\hsize]{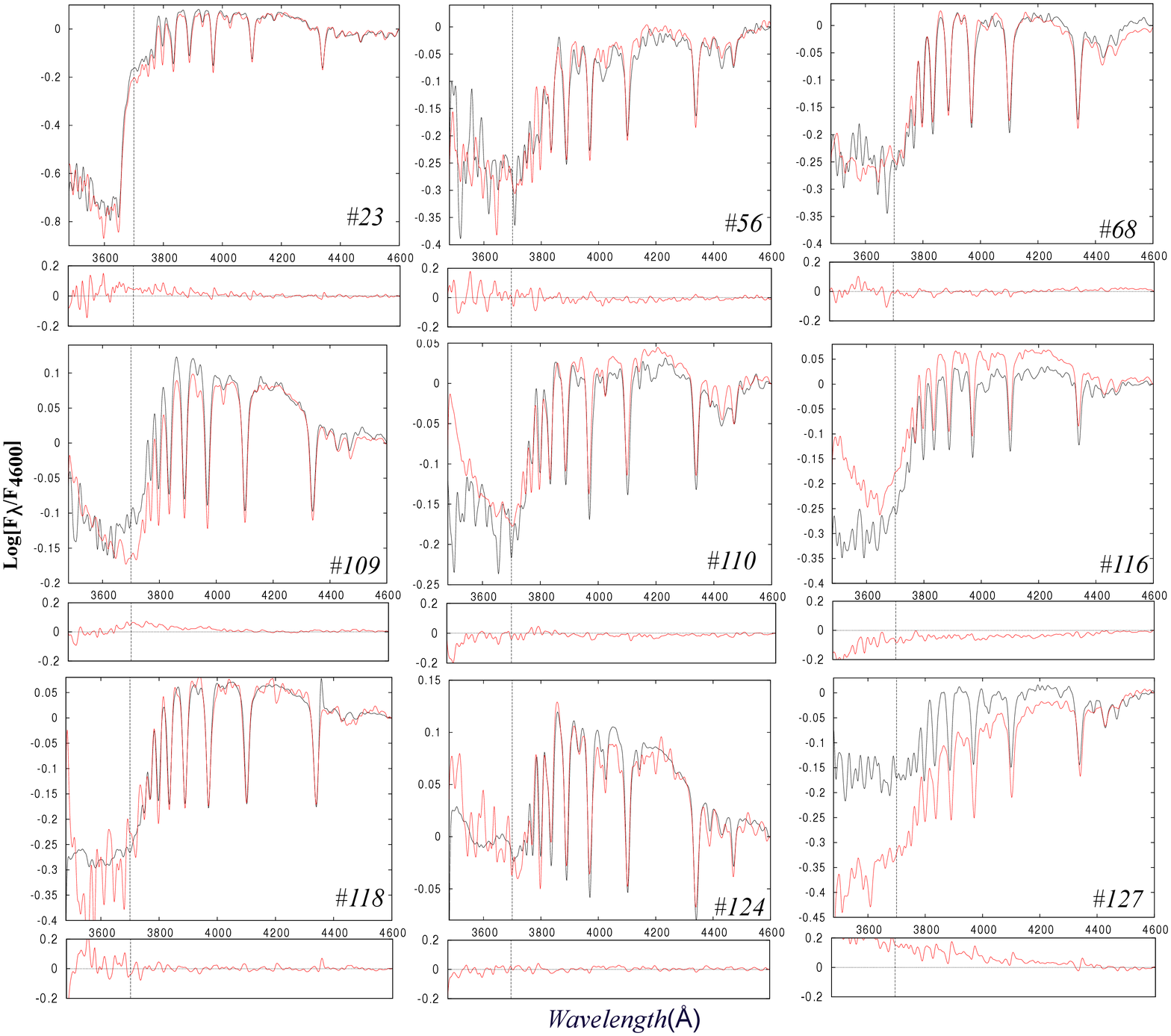}
   \caption{Stars with two or more spectra. For each object the upper panel represents the overplot of the spectra and the lower panel the difference between the two spectra.}
              \label{twospec1}%
    \end{figure*}

\begin{figure*}[b]
   \centering
   \includegraphics[angle=90, trim={1.1cm 0.0cm 0.1cm -1.5cm},clip,width=\hsize]{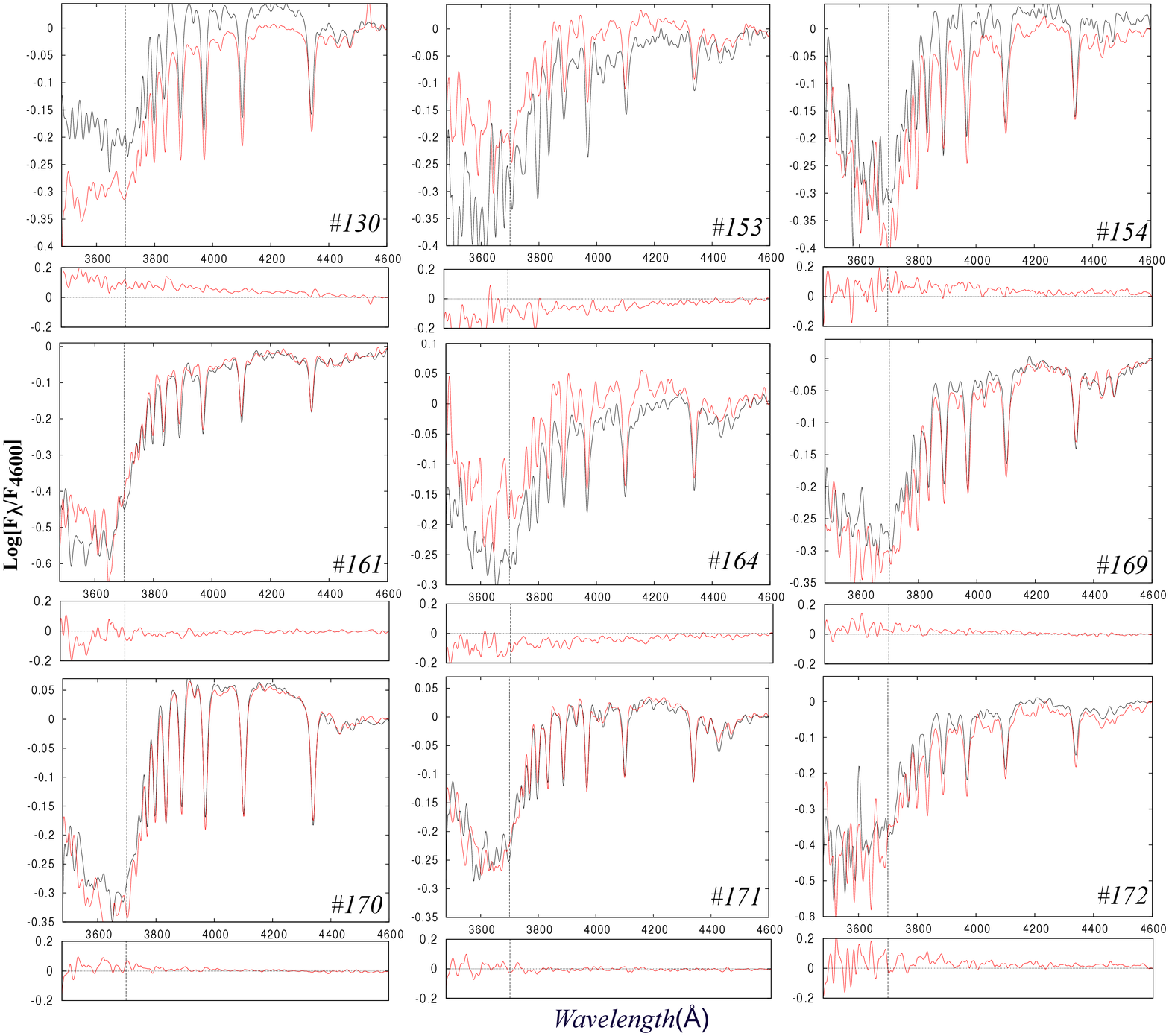}
   \caption{As Fig.~\ref{twospec1}.}
              \label{twospec2}%
    \end{figure*}

\begin{figure*}[b]
   \centering
   \includegraphics[angle=90, trim={1.1cm 0.0cm 0.1cm -1.5cm},clip,width=\hsize]{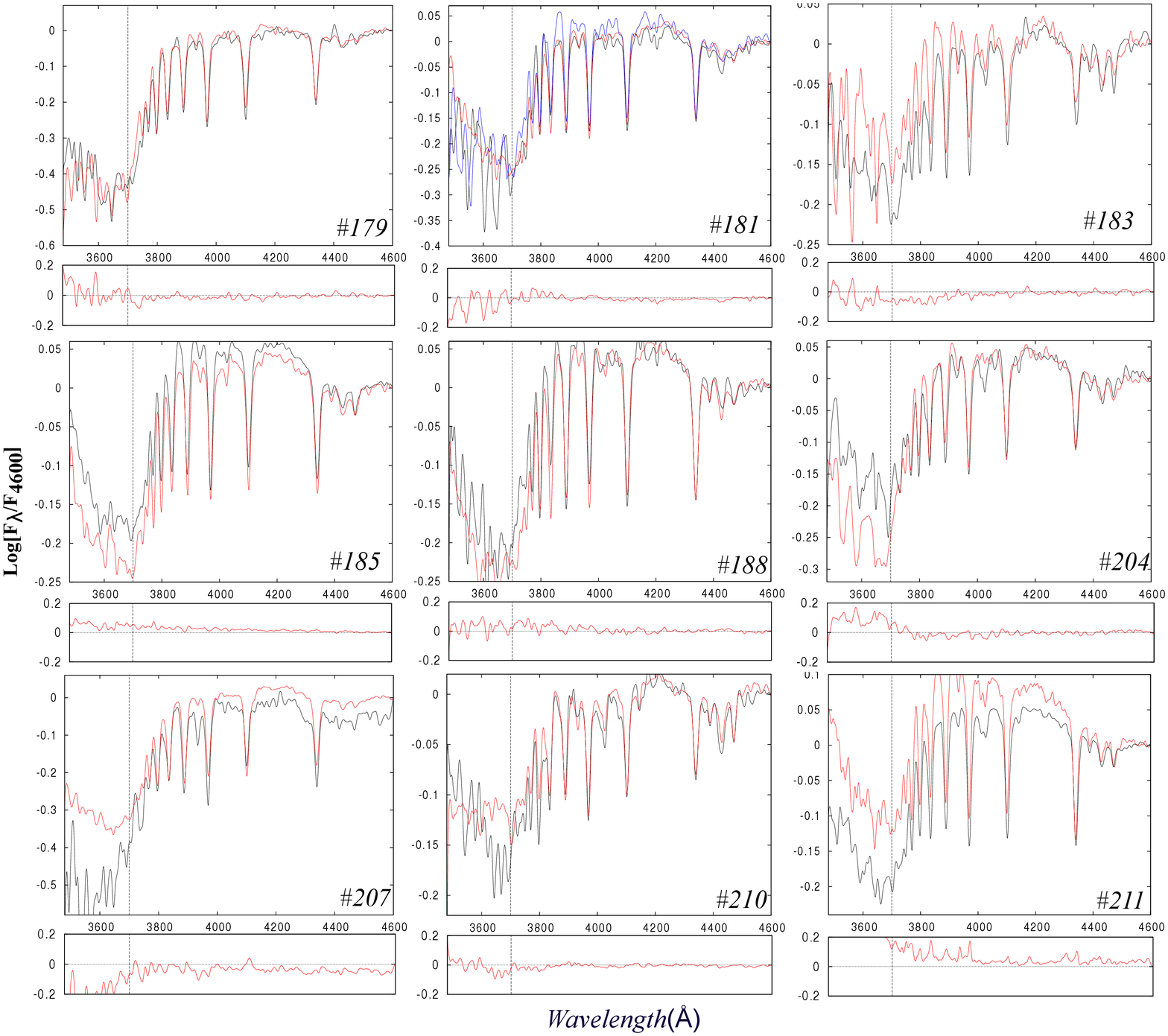}
    \caption{As Fig.~\ref{twospec1}.}
              \label{twospec3}%
    \end{figure*}

\begin{figure*}[b]
   \centering
   \includegraphics[angle=90, trim={1.1cm 0.0cm 0.1cm -1.5cm},clip,width=\hsize]{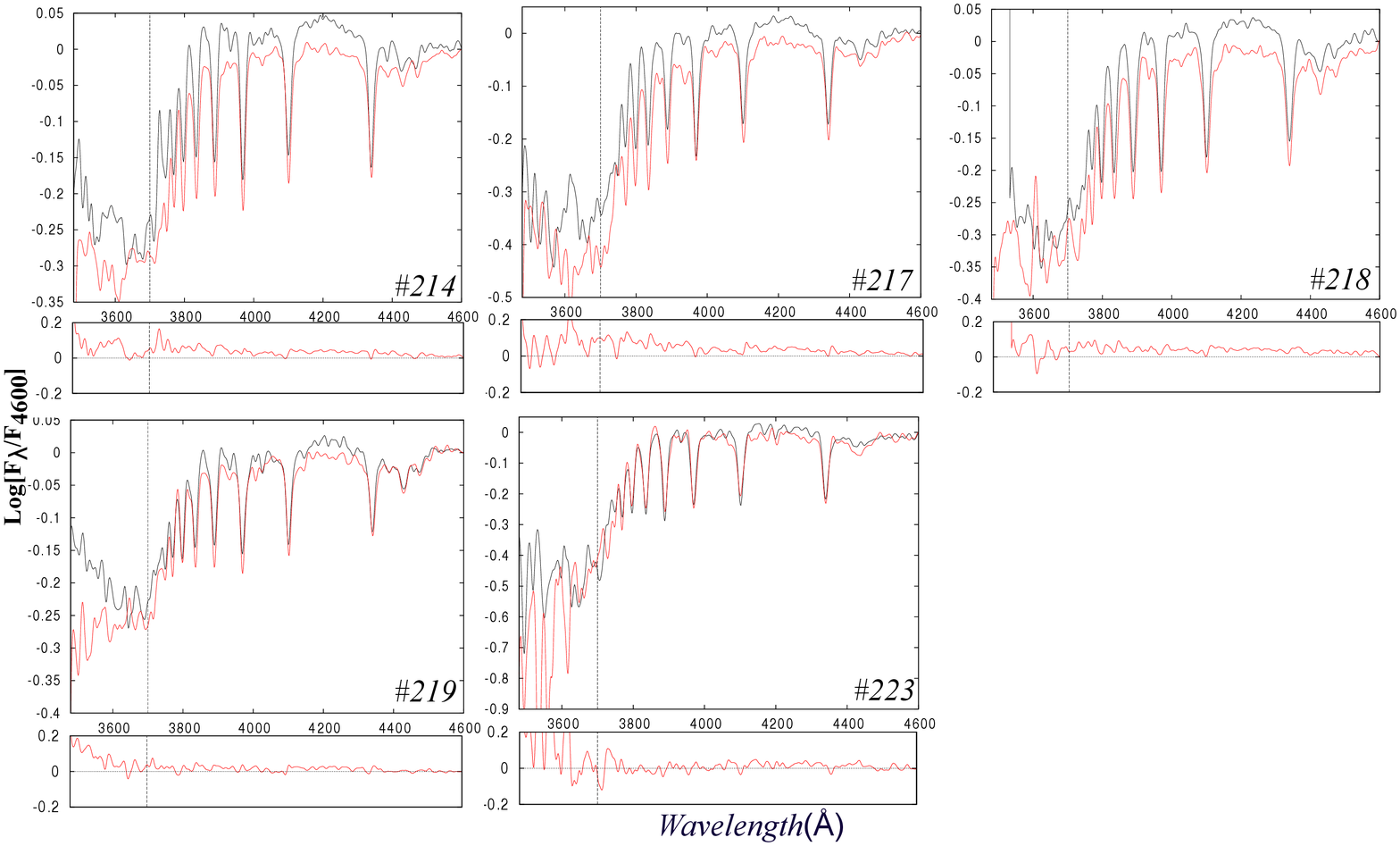}
   \caption{As Fig.~\ref{twospec1}.}
              \label{twospec4}%
    \end{figure*}

For 31 stars we have two spectra obtained at different epochs, and for one more, star 181, we have three spectra. All these spectra are presented in Figs.~\ref{twospec1} to \ref{twospec4}. For each object we present two panels. In the upper one we overplot the two- or three- flux calibrated and normalised spectra in a logarithmic scale. In the bottom panel we represent the difference between the normalised fluxes as a function of the wavelength. 

From the comparison between the spectra we can divide the object sample into three groups. Group A is composed of the stars for which the two calibrated spectra overlap, indicating that the flux calibrations applied to each spectrum are consistent. It includes 14 objects, namely stars 23, 56, 68, 110, 118, 124, 161, 170, 171, 179, 181, 210, 219 and 223. Group B is composed of the stars displaying differences between the spectra, which imply differences between the flux calibrations, but the difference is a lineal function of the wavelength, as shown in the bottom panel for each star. Group B includes 16 objects, namely stars 109, 116, 127, 130, 153, 154, 164, 169, 172, 183, 185, 188, 211, 214, 217, and 218. Finally, Group C is composed by two more objects, stars 204 and 207, which show differences in the flux calibration which are not a lineal function of the wavelength.

The differences in the flux calibration present in the stars of Group B do not have any effect on the determination of the $D$ and $\lambda_1$ parameters. If the difference between the two spectra increases linearly with the decreasing wavelength, the variation of the Paschen continuum extrapolated at 3\,700 $\AA$ exactly compensates the variation of the bottom of the Balmer discontinuity at the same wavelength, and the position of the discontinuity is not affected. The mean differences between the $D$ and $\lambda_1$ parameters for stars of Group A are $28\pm24$ dex and $7.0\pm6.0$ $\AA$ respectively, while for stars of Group B these are $31\pm22$ dex and $7.4\pm5.4$ $\AA$ respectively.

These differences do, however, affect the determination of the interstellar reddening. The reddening value is derived from the $\Phi_{b}$ parameter, which measures the slope of the Paschen continuum. This slope is sensitive to differences in the relative flux calibration. The mean difference in the determination of $E(B-V)$ for stars in the Group A is $0.06\pm0.04$ mag., while for stars in Group B it amounts to $0.15\pm0.10$ mag. This last figure is consistent with the standard deviation obtained when comparing our $E(B-V)$ values with values in the literature for the same stars presented in Sect. 4, and can be considered as the mean error of our $E(B-V)$ determination.

For stars in Group C the flux calibrations are not consistent. This represents up to four spectra out of 65 analysed in this appendix. In addition, it should be noted that the differences between the calibrations are small, and translate into mean differences of the parameters determined from them well within 3$\sigma$ of the mean errors considered in this paper.

\end{document}